\documentclass[12pt]{article}%
\usepackage[nosort]{cite}
\usepackage[usenames, dvipsnames]{xcolor}
\usepackage{graphicx}
\usepackage{multicol}
\usepackage{amsfonts}
\usepackage{amssymb}
\usepackage{amsmath}
\usepackage{heck}
\usepackage{afterpage}
\usepackage{setspace}
\usepackage{verbatim}
\usepackage{longtable}
\usepackage{float}
\usepackage{subcaption}
\usepackage{epsfig}
\usepackage{enumerate}
\usepackage{epstopdf}
\usepackage[enableskew, vcentermath]{youngtab}
\usepackage{adjustbox}
\usepackage{multirow}
\usepackage{tikz}
\usepackage[margin=1in]{geometry}
\usepackage{titletoc}
\usepackage{hyperref}
\usepackage{mathtools}%
\usepackage{tikz-cd}
\providecommand{\U}[1]{\protect\rule{.1in}{.1in}}
\pdfoutput=1
\newsavebox{\mysavebox}

\hypersetup{colorlinks,citecolor=black,filecolor=black,linkcolor=black,urlcolor=black}
\usetikzlibrary{decorations.markings}

\numberwithin{equation}{section}

\usetikzlibrary{chains}
\allowdisplaybreaks
\tikzset{node distance=2em, ch/.style={circle,draw,on chain,inner sep=2pt},chj/.style={ch,join},every path/.style={shorten >=4pt,shorten <=4pt},line width=1pt,baseline=-1ex}

\newcommand{\ba}{\begin{eqnarray}}
\newcommand{\ea}{\end{eqnarray}}

\newcommand{\be}{\begin{equation}}
\newcommand{\ee}{\end{equation}}

\tikzstyle{startstop} = [rectangle, rounded corners, minimum width=3cm, minimum height=1cm,text centered, draw=black, fill=blue!10]
\tikzstyle{startstop} = [rectangle, rounded corners, minimum width=3cm, minimum height=1cm,text centered, draw=black, fill=blue!10]
\tikzstyle{io} = [trapezium, trapezium left angle=70, trapezium right angle=110, minimum width=3cm, minimum height=1cm, text centered, draw=black, fill=blue!30]
\tikzstyle{process} = [rectangle, minimum width=3cm, minimum height=1cm, text centered, draw=black, fill=orange!30]
\tikzstyle{decision} = [diamond, minimum width=3cm, minimum height=1cm, text centered, draw=black, fill=green!30]
\tikzstyle{arrow} = [thick,->,>=stealth]
\tikzset{->-/.style={decoration={
  markings,
  mark=at position #1 with {\arrow[scale=2.4]{>}}},postaction={decorate}}}
\makeatletter \@addtoreset{equation}{section} \makeatother

\begin{document}

\vspace*{-2cm}
\begin{flushright}
{\tt UPR-1294-T}\\
\end{flushright}

\date{November 2018}

\title{F-theory and Dark Energy}

\institution{PENN}{\centerline{Department of Physics and Astronomy, University of Pennsylvania, Philadelphia, PA 19104, USA}}

\authors{Jonathan J. Heckman\footnote{e-mail: {\tt jheckman@sas.upenn.edu}},
Craig Lawrie\footnote{e-mail: {\tt craig.lawrie1729@gmail.com}},\\[4mm]
Ling Lin\footnote{e-mail: {\tt lling@physics.upenn.edu}},
and Gianluca Zoccarato\footnote{e-mail: {\tt gzoc@sas.upenn.edu}}}

\abstract{Motivated by its potential use as a starting point for solving
various cosmological constant problems, we study F-theory
compactified on the warped product $\mathbb{R}_{\text{time}} \times S^3
\times Y_{8}$ where $Y_{8}$ is a $Spin(7)$ manifold, and the $S^3$ factor
is the target space of an $SU(2)$ Wess--Zumino--Witten (WZW) model at level $N$.
Reduction to M-theory exploits the abelian duality of this WZW
model to an $S^3 / \mathbb{Z}_N$ orbifold. In the large $N$ limit,
the untwisted sector is captured by 11D supergravity.
The local dynamics of intersecting 7-branes in the $Spin(7)$ geometry
is controlled by a Donaldson--Witten twisted gauge theory coupled to defects.
At late times, the system is governed by a 1D quantum mechanics system
with a ground state annihilated by two real supercharges, which in four dimensions would appear as
``$\mathcal{N} = 1/2$ supersymmetry'' on a curved background. This leads to a
cancellation of zero point energies in the 4D field theory but a split mass
spectrum for superpartners of order $\Delta m_\text{4D} \sim \sqrt{M_\text{IR} M_\text{UV}}$
specified by the IR and UV cutoffs of the model. This is suggestively
close to the TeV scale in some scenarios. The classical
4D geometry has an intrinsic instability which can produce either a collapsing or expanding
Universe, the latter providing a promising starting
point for a number of cosmological scenarios. The resulting 1D
quantum mechanics in the time direction also
provides an appealing starting point for a more
detailed study of quantum cosmology.}

\maketitle

\setcounter{tocdepth}{2}

\tableofcontents


\newpage

\section{Introduction \label{sec:INTRO}}

One of the major obstacles in connecting string theory with the real world is
the absence of reliable examples which include explicit time dependence,
especially cosmological singularities. Part of the issue is that the most well
developed techniques in string compactification all rely on supersymmetry.

A related point is that the size of the cosmological
constant (thankfully)\ leads to a Universe much larger than the Planck scale.
This is puzzling because in a non-supersymmetric field theory the generic expectation
is to have a Planck scale contribution to the vacuum energy density. In a supersymmetric theory,
the zero point energy cancels up to terms set by the supersymmetry breaking scale, which is
still very problematic. This has led many to abandon
a dynamical explanation, instead invoking anthropics
(see e.g. \cite{Weinberg:1987dv, Bousso:2000xa}).

These issues have become particularly pressing in light of reference \cite{Obied:2018sgi} which
has made the conjectural claim that string theory may not admit pure de Sitter solutions.
Strictly speaking, we do not actually live in de Sitter space, though
it would appear to be a good first approximation of our Universe to include a dark energy sector (be it constant or dynamical)
of some sort. For earlier proposed constructions of de Sitter space in string theory, see references
\cite{Maloney:2002rr, Kachru:2003aw, Balasubramanian:2005zx,
Westphal:2006tn, Dong:2010pm, Rummel:2011cd, Blaback:2013fca, Cicoli:2013cha, Cicoli:2015ylx}, and for
a discussion of various constraints on realizing de Sitter vacua in string theory,
see for example \cite{Maldacena:2000mw, Townsend:2003qv, Hertzberg:2007wc,
Covi:2008ea, Caviezel:2008tf, Caviezel:2009tu, deCarlos:2009fq, Wrase:2010ew, Shiu:2011zt,
Green:2011cn, Gautason:2012tb, Bena:2014jaa, Kutasov:2015eba, Quigley:2015jia, Dasgupta:2014pma,
Junghans:2016abx, Junghans:2016uvg, Andriot:2016xvq, Moritz:2017xto, Sethi:2017phn, Andriot:2017jhf, Danielsson:2018ztv}.
For additional followups to reference \cite{Obied:2018sgi}, see references
\cite{Agrawal:2018mkd,Agrawal:2018own,Andriot:2018wzk,Branchina:2018xdh,Dvali:2018fqu,Banerjee:2018qey,
Aalsma:2018pll,Achucarro:2018vey,Garg:2018reu,Lehners:2018vgi,Kehagias:2018uem,Dias:2018ngv,Denef:2018etk,
Colgain:2018wgk,Roupec:2018mbn,Andriot:2018ept,Ghosh:2018fbx,Matsui:2018bsy,Ben-Dayan:2018mhe,Chiang:2018jdg,
Heisenberg:2018yae,Damian:2018tlf,Conlon:2018eyr,Kinney:2018nny,Dasgupta:2018rtp,Cicoli:2018kdo,Kachru:2018aqn,
Akrami:2018ylq,Nakai:2018hhf,Heisenberg:2018rdu,Murayama:2018lie,Marsh:2018kub,Brahma:2018hrd,Choi:2018rze,Das:2018hqy,
Danielsson:2018qpa,Wang:2018duq,Brandenberger:2018wbg,DAmico:2018mnx,Han:2018yrk,Moritz:2018ani,Bena:2018fqc,
Brandenberger:2018xnf,Dimopoulos:2018upl,Ellis:2018xdr,Lin:2018kjm,Hamaguchi:2018vtv,Kawasaki:2018daf,Motaharfar:2018zyb,
Odintsov:2018zai,Ashoorioon:2018sqb,Antoniadis:2018ngr,Das:2018rpg,Ooguri:2018wrx,Wang:2018kly,Fukuda:2018haz,Buratti:2018onj,
Hebecker:2018vxz,Gautason:2018gln,Olguin-Tejo:2018pfq,Garg:2018zdg,Dvali:2018jhn,Park:2018fuj,Blaback:2018hdo,Schimmrigk:2018gch,
Lin:2018rnx,Kim:2018mfv}.

Whatever the outcome of this debate may be, it seems worthwhile to seek out stringy backgrounds where
the size of the Universe is stable against quantum mechanical
collapse. As noted in \cite{Witten:1994cga, Witten:1995rz},
at least some of these issues can be addressed in
three-dimensional spacetimes. The key point is that in three or fewer
dimensions the ground state may
still be supersymmetric, but all excitations above the vacuum can have
a split mass spectrum. Perturbations away from this scenario would
then be capable of generating a small value for the cosmological constant.
Explicit three-dimensional vacua include compactification of
M-theory on a $Spin(7)$ manifold \cite{Becker:2000jc, Gukov:2001hf, Acharya:2002vs}.
This leaves intact two real supercharges, yielding a 3D $\mathcal{N}=1$ vacuum.

Of course, this mechanism appears to be rather special to three dimensions, and though
details are sorely lacking, it has been clear for some time that F-theory is the proper framework for seeking out a 4D lift
\cite{Vafa:1996xn}. One of the basic issues is that in 4D flat space, there is no unitary theory
with two real supercharges:\ It would be an ``$\mathcal{N}=1/2$ supersymmetric'' theory which is clearly problematic.
Previous attempts to circumvent this issue include uplifting the 3D $\mathcal{N} = 1$ vacua of M-theory on a $Spin(7)$ manifold
to F-theory reduced on a finite interval \cite{Bonetti:2013fma, Bonetti:2013nka}.

Our aim in this work will be to provide a phenomenologically viable
way to address these issues. The main thing we give up from the start
is the requirement that any excitation above the ground state be supersymmetric. We will, however,
require that the ground state is still annihilated by two supercharges. To get there, we
shall consider spacetimes with topology $\mathbb{R}_{\text{time}}\times S^{3}$ of the same sort which appears in
Einstein's static Universe \cite{Einstein:1917}. The full F-theory
background will then be described by a warped product of the form
$\mathbb{R}_{\text{time}} \times S^3 \times Y_8$ with $Y_8$ a $Spin(7)$ manifold.\footnote{Throughout
this paper we adopt a maximally flexible notion of ``$Spin(7)$.'' All we
shall require is a Killing spinor which would preserve two pseudo-real supercharges.
In particular, we only demand that the structure group of the tangent bundle be $Spin(7)$.}
Note that our proposal is in some sense
related to earlier ones in \cite{Bonetti:2013fma,Bonetti:2013nka}
because we can view an $S^3$ as a $T^2$ fibration over a finite interval.

We argue that the ground state of the resulting 4D system is still annihilated
by two supercharges, but for all finite energy excitations, supersymmetry is broken.
This follows directly from an analysis of the Killing spinor equations. In particular, the ground
state remains supersymmetric because in signature $(2,2)$ (the analytic continuation of our spacetime), two real supercharges
can be preserved. This continuation serves to define the ground state and we shall (by extension of related terminology)
refer to it as the ``Bunch-Davies'' vacuum. Even so, analytic continuation tells
us little about the spectrum of finite energy excitations in signature $(3,1)$.

Interestingly enough, although it is by itself not a good description of the real world,
the Einstein static Universe is the starting point for various
cosmological scenarios, such as references \cite{Ellis:2002we, Barrow:2003ni, Ellis:2003qz}.
A further remark is that although this involves a closed FRW Universe, present observational constraints
do not exclude this possibility. We shall leave a more detailed phenomenological analysis for future work.

The background with 4D spacetime $\mathbb{R}_{\text{time}} \times S^3$ is amenable to an exact analysis
because an $S^{3}$ is the target space of an $SU(2)$ Wess--Zumino--Witten (WZW) model \cite{Witten:1983ar}
at large level \cite{Gepner:1986wi} (for a
wonderfully concise review see reference \cite{Schulz:2011ye}).\footnote{In fact, there is a whole branch of stringy
cosmological models based on using exactly solvable WZW models. Here, we do not demand all of the
time dependence be captured by a worldsheet CFT, just a metastable starting point.
For some early work in this direction see for example
references \cite{Antoniadis:1988vi, Bars:1989ph, Bars:1990rb, Ginsparg:1992af, Nappi:1992kv}.}
In the target space picture, the size of the $S^{3}$ is controlled by the number
of units of Neveu--Schwarz (NS) three-form flux threading the geometry.
The 10D\ spacetime then takes the
form of a warped product $\mathbb{R}_{\text{time}}\times S^{3}\times B_{6}$,
where $B_{6}$ is a six-manifold which is the base of a torus-fibered
eight-manifold $Y_{8}$.

Indeed, there has recently been substantial progress in the explicit construction of both
local and global $Spin(7)$ manifolds, and there does not appear to be any a priori obstruction to
demanding the additional condition of a $T^2$ fibration (as required by F-theory).
For example, orbifolds / involutions of $T^8$ \cite{Joyce1996, Joyce:1999, Clancy:2011, Lee:2014wma}, the
twisted connected sum construction (see \cite{MR2024648}) of \cite{Braun:2018joh} and local models based on non-holomorphic
K3 fibrations over four-manifolds \cite{Gukov:2001hf} (see also \cite{bryant1989,Gibbons:1989er,Cvetic:2001pga, Cvetic:2001ye, Kanno:2001xh, Cvetic:2001zx}) all appear to be compatible with the existence of a torus fibration.

But one generic feature of these constructions is the absence of a \textit{holomorphic} profile for the axio-dilaton.
This means that when attempting to solve the Einstein field equations, the dilaton equations of motion will necessarily need
to be sourced by contributions other than just 7-branes. Our proposal is that we need only allow for $H$-flux in the
4D directions of the spacetime to alleviate this issue. From the perspective of a 4D observer,
this looks like a peculiar combination of flux through an $S^3$ and a positive cosmological constant (the remnant
of having a non-holomorphic axio-dilaton profile). We stress that the two contributions cannot be disentangled.

\begin{figure}[t!]
\begin{center}
\includegraphics[width = \textwidth]{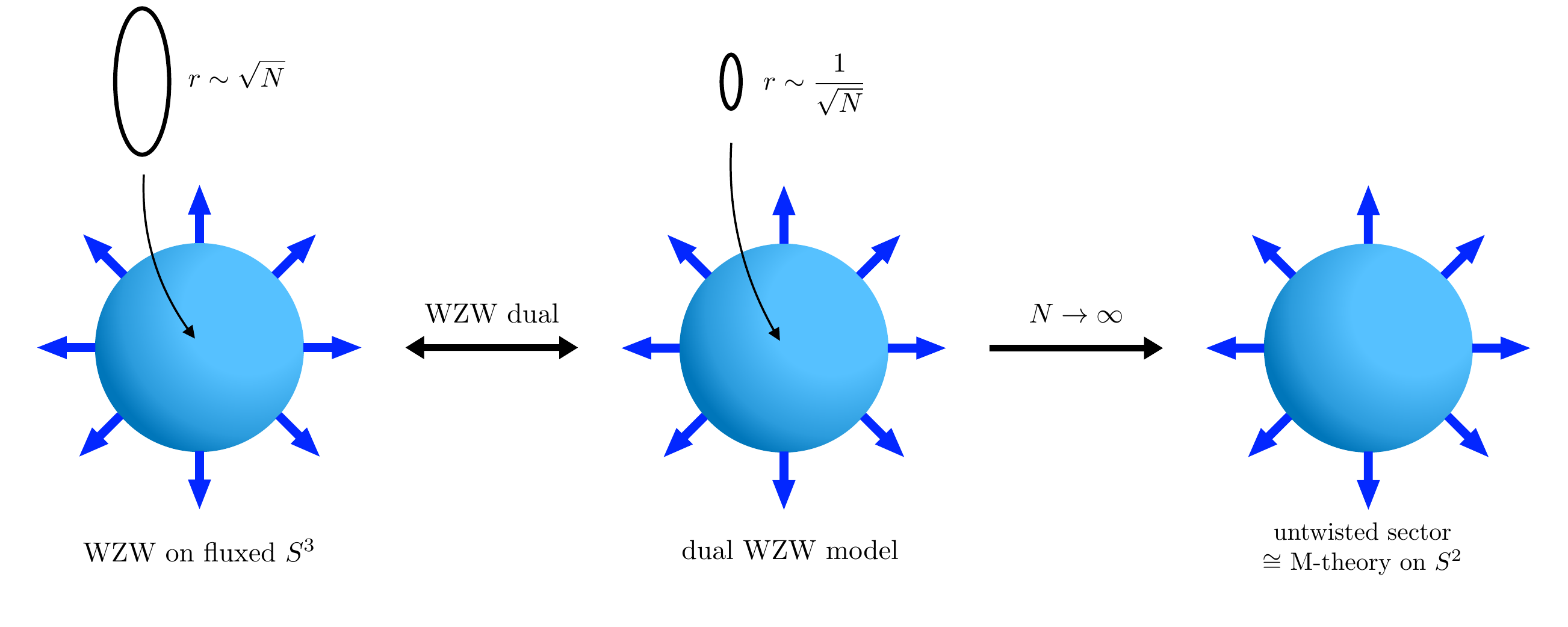}
\end{center}
\caption{
	The WZW-interpretation of F-theory on $S^3 \times Y_{8}$ allows for a T-dual description in terms of the Hopf fiber of the $S^3$ with $N$ units of flux (left and middle).
	The large $N$ limit projects onto the untwisted sector which corresponds to M-theory on $S^2 \times Y_{8}$ (right).
}
\label{fig:Untwisted}
\end{figure}

The present construction also provides a candidate M-theory description of
some of the resulting physics. Recall that in the context of Calabi--Yau
compactification of F-theory, reduction on a further circle is described (up
to possible twists on the circle used for dimensional reduction)
by M-theory on the same Calabi--Yau. In the present
construction, the role of this circle is played by the
$S^{1}$ in the\ Hopf fibration of the $S^{3}$. Indeed, using the fact that at
level $N=pq$, the $SU(2)/\mathbb{Z}_{p}$ and $SU(2)/\mathbb{Z}_{q}$ WZW\ models are equivalent \cite{Gaberdiel:1995mx, Maldacena:2001ky}, we learn that the large $N$ limit of our
geometry has an alternative description in terms of a target space $SU(2)/\mathbb{Z}_{N} \stackrel{N \rightarrow \infty}{\simeq} S^{2}$. To maintain unitarity in the model it is of course essential to include all of
the twisted sectors, but projecting onto just the untwisted sector, we reach an 11D
background for M-theory on the warped product
$\mathbb{R}_{\text{time}}\times S^{2}\times Y_{8}$, with $Y_{8}$ a
$Spin(7)$ manifold. In this geometry, there is a four-form flux threading the
$S^{2}$ as well as the torus class of the fiber of $Y_{8}$.\ We also see
that this M-theory description becomes inadequate at sufficiently high
energies because to retain a unitary description of physics, one must also
include the other twisted sectors of the WZW model. See figure \ref{fig:Untwisted}
for a depiction.

Regardless of whether we are dealing
with M-theory or F-theory on a given background, we expect some details to be universal in both treatments,
especially in the long time limit captured by the 1D quantum mechanics on $\mathbb{R}_{\text{time}}$.
With this in mind, we first analyze the case of M-theory compactified on a local $Spin(7)$ manifold described locally as a non-holomorphic K3 fibration over a four-manifold. Geometrically, this is quite similar in spirit to the analysis of \cite{Acharya:2001gy} for local $G_2$ spaces
(see also \cite{Pantev:2009de}). Here, we develop a similar
framework for 7D super-Yang--Mills theory wrapped on a four-manifold. In particular, we find that this is compatible with 3D $\mathcal{N} = 1$
supersymmetry via the Donaldson--Witten twist \cite{Witten:1988ze}. This system
can be generalized to include matter fields localized on curves in the
four-manifold, as well as triple intersections of curves at points. All of this is quite similar to the prescription used in F-theory GUT model
building \cite{Beasley:2008dc, Donagi:2008ca} (for reviews see \cite{Heckman:2010bq, Weigand:2010wm}),
so it is perhaps not surprising that it carries over to this case as well.

Treating the 3D effective theory on $\mathbb{R}_{\text{time}} \times S^2$ as part of an untwisted sector of a full orbifold theory,
we can also lift this description back to a 4D effective theory on $\mathbb{R}_{\text{time}} \times S^3$. In F-theory,
we are accustomed to viewing the discriminant locus of a holomorphic elliptically fibered Calabi--Yau space as the locus wrapped by 7-branes.
Since we do not have a holomorphic K3 fibration, however, this is clearly more delicate. Our interpretation is that there is
indeed a 7-brane gauge theory since locally, we can always model the K3-fibration holomorphically. However, since this fibration is globally non-holomorphic (even in non-compact models), there is always an anti-holomorphic image $\widetilde{7}$-brane which also participates in the local model. The mode content for the holomorphic brane with its anti-holomorphic image leads to the full mode content of the 4D system, and also correctly reproduces the 3D untwisted sector limit as described by 7D super-Yang--Mills theory wrapped on a four-manifold.

This analysis also indicates that at least locally, the $Spin(7)$ background
can be viewed as the backreacted geometry obtained from F-theory on a
Calabi--Yau fourfold background in the presence of additional NS5-branes.
Though we leave a full analysis of this for future work, we can already see
that this leads, for example, to a 4D background with a dilaton gradient in
the internal dimensions. Note that if we explicitly break 4D Lorentz symmetry,
there is no issue with preserving just two real supercharges.

We also study small fluctuations of the closed string sector using the 1D quantum mechanics.
In this way, we can also identify the quantum dynamics of the scale factor $a(t)$ which appears as
a \textquotedblleft volume modulus\textquotedblright\ of the FRW\ class of
metrics:%
\begin{equation}
  ds^{2}=-dt^{2}+a(t)^{2}d\Omega_{S^{3}}^{2} \,.
\end{equation}
Indeed, a notorious issue in string compactification is that at least some moduli are
prone to runaway behavior. Here, we see a pleasant outcome of this generic
behavior:\ Possible runaway behavior of our modulus is a welcome feature and
indicates a model with an ever expanding Universe!

Finally, we also address the sense in which we expect our 4D theory to have supersymmetry.
The construction presented in this paper leads, in the 3D M-theory limit, to a model where the bosons and fermions
have a split mass spectrum, induced by gravitational interactions. This is a peculiarity of 3D models, and the
``dark fantasy'' of references \cite{Witten:1994cga, Witten:1995rz} is that this may lead to a sizable mass splitting
in any candidate 4D model which implements these features. We find that the 4D mass splitting in our model is given
by the geometric mean of the IR and UV cutoffs:
\begin{equation}
  \Delta m_\text{4D} \sim \sqrt{M_\text{IR} M_\text{UV}} \,.
\end{equation}
We remark that using the observed energy density from dark energy and
the Planck scale to respectively set $M_\text{IR}$ and $M_\text{UV}$, this
yields a rough estimate for $\Delta m_\text{4D} $ on the order of the TeV scale.

The rest of this paper is organized as follows. Section \ref{sec:REVIEW} motivates the
study of F-theory on $Spin(7)$ backgrounds by emphasizing that there appears
to be no obstruction to generating geometries with a $T^2$ fibration. Section \ref{sec:MFMF}
presents the general proposal that F-theory on $Spin(7)$
backgrounds really requires us to work on a curved 4D spacetime. The M-theory limit is then
obtained by an abelian T-duality and projection onto the untwisted sector.
Following this, in section \ref{sec:KILLING} we consider the resulting Killing spinor equations in the presence of flux.
In section \ref{sec:MTHEORY} we turn to the 3D physics of local models in M-theory on such backgrounds, and section \ref{sec:FTHEORY}
presents the lift to four dimensions. Section \ref{sec:CLOSED} turns to some universal aspects of the closed string sector,
in particular the dynamics of the scale factor and its interpretation in the associated 1D supersymmetric quantum mechanics.
In section \ref{sec:FANTASY} we estimate the size of mass splittings for bosons and fermions in the 4D model.
Section \ref{sec:CONC} presents our conclusions. Some additional review items and additional technical details are
deferred to the Appendices.

\section{Torus Fibrations in \boldmath{$Spin(7)$} Manifolds \label{sec:REVIEW}}

Our aim in this section is to motivate the study of F-theory on $Spin(7)$ backgrounds.
This will lead to a sharp puzzle because the resulting 4D model in flat space
would appear to have too few supercharges to support a consistent interpretation.
We resolve this puzzle in sections \ref{sec:MFMF} and \ref{sec:KILLING}.

The general paradigm here is essentially the same as in other contexts:
F-theory \cite{Vafa:1996xn, Morrison:1996na, Morrison:1996pp}
geometrizes the Type IIB $SL(2,\mathbb{Z})$
duality acting on the axio-dilaton $\tau_\text{IIB}$ in terms of the complex structure
of an elliptic curve. The profile of this axio-dilaton over the 10D spacetime then generates a 12D geometry given by a
$T^2$ fibration over the base. Much of the focus in the literature has centered on models of the form:
\begin{equation}
  \mathbb{R}^{D-1 , 1} \times \text{CY}_{n} \,,
\end{equation}
with CY$_n$ an elliptically fibered Calabi--Yau manifold of complex dimension $n$ such that $D + 2(n-1) = 10$. Indeed,
the base of the elliptic fibration is a K\"ahler manifold of complex dimension $n-1$. More generally, it is not even necessary to
have a holomorphic section for the elliptic fibration and this broader class
of models is often referred to as a genus one fibration \cite{Braun:2014oya,Morrison:2014era,
Anderson:2014yva,Mayrhofer:2014haa,Cvetic:2015moa}
(see \cite{Weigand:2018rez,Cvetic:2018bni} for reviews).

A pleasant feature of F-theory on a Calabi--Yau is that
the internal components of the Type IIB Einstein field equations with holomorphically varying axio-dilaton
is automatically solved. Additionally, the Calabi--Yau condition ensures that
minimal flat space supersymmetry is achieved in the uncompactified directions.

Even so, there is a priori no reason not to consider more general F-theory backgrounds. In particular, there has, for a long time,
been a conceptual puzzle of how to make sense of F-theory on a $Spin(7)$ background since it would appear to preserve half the supersymmetry
of the minimal case obtained from a Calabi--Yau fourfold. One thing we must give up from the start is
the assumption that we have a holomorphically varying axio-dilaton. This will play an important role in the physical
proposal of section \ref{sec:MFMF}.

To begin, we shall recall some of the key features of $Spin(7)$ manifolds. We label these as eight-dimensional manifolds $Y_{8}$.
First of all, $Y_{8}$ is an orientable Riemannian manifold with a Ricci-flat metric $g$ whose holonomy is a subgroup of $Spin(7)$.
Such a manifold comes equipped with a distinguished closed, self-dual four-form $\Omega_{(4)}$, the so-called Cayley-form, which defines a calibration on $Y_{8}$.
The Cayley-form is stabilized under the $Spin(7)$ holonomy.

Explicit constructions of such manifolds have only been presented fairly recently.
For global examples, one typically relies on algebraic methods and certain underlying holomorphic structures.
Indeed, the first compact $Spin(7)$ spaces were constructed by Joyce as quotients of $T^8$ by discrete isometries $\Gamma$ \cite{Joyce1996} with only isolated fixed points.
Of course, in such a case, the quotient will locally still look like an eight-torus, so it is plausible that the total space will retain, most likely multiple, torus fibrations, which however will in general not be holomorphic.
A very concrete model of this type has also been presented in \cite{Lee:2014wma}.

To generalize the quotient construction, Joyce realized in \cite{Joyce:1999nk} that one can also obtain $Spin(7)$ holonomy via anti-holomorphic involutions $\sigma$ of a Calabi--Yau fourfold $Z$.
Essentially, this is because on a CY$_4$ with holomorphic $(4,0)$-form $\Omega_{(4,0)}$ and K\"ahler form $J$, there is a four-form:
\begin{equation}\label{eq:Cayley_fourform_from_CY4}
	\Omega_{(4)} := \text{Re} \left( e^{i \theta} \, \Omega_{(4,0)} \right) + \frac{1}{2} \, J \wedge J \, ,
\end{equation}
which is stabilized by a $Spin(7)$ subgroup of the general $SO(8)$ holonomy. In section \ref{sec:MTHEORY} we
will interpret this as a choice of 3D $\mathcal{N} = 1$ subalgebra of $\mathcal{N} = 2$ supersymmetry.
Now, if the anti-holomorphic involution $\sigma: Z \rightarrow Z$ induces the action $\sigma^* (\Omega_{(4,0)}) = e^{2 i \theta} \, \overline{\Omega}_{(0,4)}$, then $Y / \sigma$ is a $Spin(7)$ manifold with Cayley-form given by \eqref{eq:Cayley_fourform_from_CY4}.
Such a quotient construction can now be easily applied to elliptically or K3 fibered CY$_4$'s, in which case the quotient will inherit the generic fiber from the CY$_4$.
That is, locally, one still expects to have a fibration of tori or K3 surfaces over a base.
Globally however, the fibration will no longer be holomorphic.

More recently, other types of compact models have been constructed as so-called generalized connected sums (GCS) \cite{Braun:2018joh}.
In simple terms, a GCS-$Spin(7)$ manifold can be obtained from two building blocks, one of which is simply an open, Calabi--Yau fourfold with asymptotic region $Z_+ \rightarrow \text{CY}_3 \times S^1 \times \text{interval}$.
The other one is an open $G_2$ manifold $N_{7}$ over a circle, i.e., $Z_- \cong N_{7} \times S^1$, where $N_{7}$ asymptotes $\text{CY}_3 \times \text{interval}$ with the same Calabi--Yau threefold CY$_3$ as in the asymptotic region of $Z_+$.
The gluing along the asymptotic region produces a $Spin(7)$ manifold $Y_{8}$.
We summarize some of the details laid out in \cite{Braun:2018joh} in Appendix \ref{app:spin7_construction}.
For the purpose of this and the subsequent sections, we note that it is again easy to at least locally identify torus (as well as K3) fibrations.
In fact, in the spirit of the so-called twisted connected sum (TCS)
construction of $G_2$ manifolds \cite{MR2024648,MR3109862,Corti:2012kd},
$N_{7}$ will in general have an underlying K3 fibration, which can be matched
with a suitable K3 fibration on the CY$_3$ on the $Z_+$ component.
Because of the gluing, for both the TCS-$G_2$ and the GCS-$Spin(7)$, the resulting fibration will generically not be holomorphic.

\subsection{Local K3 Fibrations}

While we have given some qualitative arguments that global constructions can accommodate torus fibrations as well as
K3 fibrations in a natural way, such structures have been utilized explicitly in local models \cite{bryant1989, Gibbons:1989er,Cvetic:2001pga,Cvetic:2001ye,Kanno:2001xh,Cvetic:2001zx,Gukov:2001hf, Gukov:2002zg}.
The motivation in these papers was to obtain concrete forms of complete $Spin(7)$ metrics.
In addition, it is possible to engineer certain types of singularities.
Specifically, these local $Spin(7)$ spaces are cones over coset spaces $SO(5)/SO(3)$ or $SU(3)/U(1)$, such that the conical singularities can be viewed as a singular limit of an asymptotically locally Euclidean (ALE) space.
Equivalently, the total space can be viewed as a Taub--NUT fibration over a four-cycle $S$.
Of course, this is just a local description of a K3 fibration over $S$.

Rather than launch into the technical details presented in the above references, we shall
instead focus on the main physical points. First, we employ the duality between M-theory and type IIA:
\begin{equation}
	\text{M-theory on } Y = N \times S^1 \quad \Longrightarrow \quad \text{type IIA on } N\, .
\end{equation}
Moreover, in the spirit of F-theory, one can also allow for M-theory on an $S^1$-fibration $Y \rightarrow N$ and invoke the limit fiberwise, with singularities indicating the presence of D6-branes.
For the purposes of this section, let us restrict to compactifications to 3D Minkowski space, in which case $Y = Y_{8}$ is an eight-manifold and $N = N_{7}$ a seven-manifold.

For M-theory compactifications preserving ${\cal N} = 1$ SUSY in 3D flat space, we already know that $Y_{8}$ has to be $Spin(7)$.
On the other side, type IIA requires $N_{7}$ to be a $G_2$ manifold, with D6-branes wrapping co-associative four-cycles $X \subset N_{7}$ \cite{Becker:1995kb, Gauntlett:1998vk, Gomis:2001vk}.\footnote{
	That is, $X$ is a four-cycle calibrated by the four-form $\star \varphi$, where $\varphi$ is the three-form associated with the $G_2$ metric.
}
However, it is also known that the M-theory uplift of D6-branes gives rise to a local Taub--NUT space.
More precisely, for a stack of D6-branes on $X$, the dual M-theory is a purely geometric background where the local geometry is a Taub--NUT fibration over $X$.
Of course, a Taub--NUT space is itself just a local description around the singularities of a K3, i.e., it describes the patch around an ADE singularity $\mathbb{C}^2 / \Gamma$, where $\Gamma$ is a finite subgroup of $SU(2)$.
Thus, the physical intuition from M-theory / IIA duality precisely gives us the local picture of a $Spin(7)$ manifold $Y_{8}$ as a K3-fibration over a four-cycle $X$.
Note that the co-associativity of $X$ now implies that it is calibrated by the Cayley-form $\Omega_{(4)}$ on $Y_{8}$.
In general, $X$ is a four-dimensional Riemannian manifold endowed with a $Spin^c$ structure.
If we had a 3D $\mathcal{N} = 2$ theory, the local geometry would be a Calabi--Yau fourfold and $X$ would be a K\"ahler surface.
We will return to this point in section \ref{sec:specialization}.

\section{F-theory and M-theory on \boldmath{$Spin(7)$} Backgrounds}\label{sec:MFMF}

The main lesson from the previous section is that there is
no reason, a priori, to not consider non-holomorphic torus fibered $Spin(7)$ backgrounds.
From this perspective, we must ask how to make sense of the resulting F-theory backgrounds.

This presents an immediate puzzle because in flat space, F-theory on an elliptically fibered
Calabi--Yau fourfold already has minimal $\mathcal{N} = 1$ supersymmetry. A torus fibered $Spin(7)$
background would have ``$\mathcal{N} = 1/2$ supersymmetry.'' The main aim of this section is to present a
general phenomenologically viable proposal for how to make sense of such backgrounds.

M-theory on $Spin(7)$ backgrounds has no such problems. Indeed
reference \cite{Becker:2000jc} explicitly showed that M-theory can be placed on
warped products of the form $\mathbb{R}^{2,1}\times Y_{8}$ with $Y_{8}$ a
$Spin(7)$ manifold. In section \ref{sec:KILLING} we show that from the
perspective of the Killing spinor equations,
there is no obstruction to switching on a background four-form flux which threads the
spatial components of the geometry as well as some two-cycle of the internal
eight-manifold. Doing so, we should technically relax the constraint of
$Spin(7)$ metric holonomy, and only demand an internal manifold with structure
group $Spin(7)$, though in what follows we shall ignore such fine points.
In a local frame there is no obstruction to retaining 3D $\mathcal{N} = 1$ in the presence of a spatial
flux. So in principle, this construction can be broadened to consider warped products of the form
$\mathbb{R}_{\text{time}}\times M_2 \times Y_{8}$ with suitable four-form flux threading internal four-cycles
as well as a compact two-cycle in the spacetime (and an internal two-cycle).\footnote{Globally,
there are obstructions to retaining a well-defined Killing spinor, especially when this spatial factor is a round $S^2$.}

For the purposes of F-theory we must restrict to torus fibered $Spin(7)$ backgrounds. Already, there is a
potential issue because in 4D flat space, there is no theory with two real supercharges.
On the other hand, an important clue from the discussion of section \ref{sec:REVIEW} is
that the torus fibration is necessarily non-holomorphic. That means the IIB axio-dilaton
$\tau_{IIB}(z , \overline{z})$ will generically have strong internal gradients in the
internal directions. To satisfy the Einstein field equations one must then include additional
sources beyond what 7-branes can provide.

Here, we consider the simplest way to address this issue: We consider F-theory
compactified on the warped product $\mathbb{R}_{\text{time}}\times S^{3}\times
Y_{8}$, where the $S^{3}$ is really the target space of an $SU(2)$ WZW\ model
at large level. F-theory compactified on warped products to non-Minkowski
spacetimes has also been studied recently in the context of the AdS/CFT
correspondence in \cite{Couzens:2017way,Couzens:2017nnr}.

To see why this has a chance of working, recall
that the WZW\ model is defined by a 2D\ CFT\ with target space metric and
$B$-field. Treating these parameters as couplings of the 2D\ theory, one finds
that the beta functions can be consistently set to zero, yielding a conformal
fixed point \cite{Witten:1983ar}. This is a promising start for solving all of the
equations of motion of the NS\ sector. Indeed, at lowest order in
$\alpha^{\prime}$ one gets the following beta functions
(in the critical dimension):\footnote{See for
instance section 3.7 of \cite{Polchinski:1998rq}.}
\begin{align}
\beta_{\mu\nu}^{g}  &  =\alpha^{\prime}R_{\mu\nu}+2\alpha^{\prime}\nabla_{\mu
}\nabla_{\nu}\Phi-\frac{1}{4}\alpha^{\prime}H_{\mu\rho\sigma}H_{\nu}{}%
^{\rho\sigma}\,,\\
\beta_{\mu\nu}^{B}  &  =-\frac{\alpha^{\prime}}{2}\nabla^{\rho}H_{\rho\mu\nu
}+\alpha^{\prime}\nabla^{\rho}\Phi H_{\rho\mu\nu}\,,\\
\beta^{\Phi}  &  =-\frac{\alpha^{\prime}}{2}\nabla^{2}\Phi+\alpha^{\prime
}\nabla_{\mu}\Phi\nabla^{\mu}\Phi-\frac{\alpha^{\prime}}{24}H_{\mu\nu\rho
}H^{\mu\nu\rho}\,.
\end{align}

The central charge of the bosonic $SU(2)$ WZW model at level $N$ is:%
\begin{equation}
c=\frac{3N}{N+2},
\end{equation}
so in the large $N$ limit, we get $c=3$, as required to build three
macroscopically large dimensions. At finite $N$, we also observe a deficit
in the central charge:%
\begin{equation}
\delta c=3-\frac{3N}{N+2}=\frac{6}{N+2}. \label{deficit}%
\end{equation}
Indeed, to satisfy $\beta^{\Phi} = 0$ one must include a gradient of some sort
for the dilaton. This is due to two features. First, the WZW model is in dimension ``$3 - \varepsilon$''
and second, the presence of a non-zero $H$-flux means we cannot
otherwise solve for a vanishing beta function $\beta^{\Phi}$. The classic
example of this sort is the near horizon geometry of an NS5-brane \cite{Callan:1991at}.

From the perspective of a 4D observer on the $\mathbb{R}_{\text{time}} \times S^3$ factor,
the appearance of a dilaton gradient in the internal directions would look like a positive cosmological constant.
At present we do not see how to reliably disentangle this from the presence of the $H$-flux, so
this definitely would not produce a pure de Sitter solution.
We return to this point in section \ref{sec:CLOSED}.

From the perspective of the 12D F-theory geometry, we also need this $H$-flux because by assumption,
the axio-dilaton is not varying holomorphically. This means that in the $\beta^{\Phi} = 0$ equation
of motion, the internal contributions cannot be cancelled off \textit{unless} some other fields
contribute. Here we have taken the simplest possibility as sourced by NS three-form flux but it would
of course be interesting to consider more general possibilities.

Returning to the beta function equations of motion, the WZW\ ansatz then
applies so $\beta_{\mu\nu}^{B}$ is zero. The beta
function $\beta_{\mu\nu}^{g}$ is also zero because we require:%
\begin{equation}
R_{\mu\nu}=\frac{1}{4}H_{\mu\rho\sigma}H_{\nu}{}^{\rho\sigma}\,.
\end{equation}
Finally, as already mentioned, $\beta^{\Phi}$ is zero because we allow
for a non-trivial dilaton gradient in the extra dimensions.

We can also write down the explicit metric and $H$-flux which solve these
equations. As is standard, we write the unit three-sphere metric as:%
\begin{equation}
d\Omega_{(3)}^{2}=d\psi^{2}+\sin^{2}\psi\,d\theta^{2}+\sin^{2}\psi\sin
^{2}\theta\,d\phi^{2}\,,
\end{equation}
and the $H$-flux is proportional to the volume form on the unit $S^{3}%
$:%
\begin{equation}
H=\frac{Nl_{s}^{2}}{2\pi^{2}}\,\text{Vol}_{S^{3}} \, ,
\end{equation}
where Vol$_{S^3}=\sin^{2}\psi\sin\theta$
$d\psi\wedge d\theta\wedge d\phi$ is the volume form on a unit radius 3-sphere.
Our units for $l_{s}^{2}$ are chosen so that we obey the proper quantization
condition:%
\begin{equation}
\frac{1}{l_{s}^{2}}\int_{S^{3}}H=N\,.
\end{equation}
Including supersymmetry is also straightforward in this framework, a point we shortly turn to
in section \ref{sec:KILLING}. Indeed, no supersymmetry is broken by such WZW models.
In this sense, the WZW\ model is actually a \textquotedblleft
spectator.\textquotedblright\

At a more conceptual level, introducing the background $H$-flux amounts to incorporating torsion into the
spin connection. This affects left handed and right-handed spinors differently, and is captured by
the substitutions we perform for the two connections:
\begin{align}
\omega^{(L)}(H) & =  \omega + \frac{1}{2} H \, , \label{leftshift}\\
\omega^{(R)}(H) & =  \omega - \frac{1}{2} H \, . \label{rightshift}
\end{align}
The WZW model on $S^3$ automatically solves the Einstein field equations because
the curvature with respect to the shifted spin connections vanish. Indeed, an $S^3$ is parallelizable.

Another wonderful feature of the WZW model is that the CFT\ description
is essentially exact and solvable. In particular, the model enjoys an $SU(2)_{L}\times
SU(2)_{R}$ symmetry, and we can label states of the CFT as descendants of the
Virasoro primaries:%
\begin{equation}
  \left\vert j_{L},m_{L};j_{R},m_{R}\right\rangle \,,
\end{equation}
where $j_{L},j_{R}$ denote the particular representation in question, and
$m_{L},m_{R}$ denote values of the $J_{3}^{(L)}$ and $J_{3}^{(R)}$ operators.
These should be viewed as labelling the \textquotedblleft
momenta\textquotedblright\ of a state on the $S^{3}$, and are also associated
with the standard tachyonic states present in any bosonic string theory. In
the worldsheet model, the values of the angular momenta are cut off by the
conditions:
\begin{equation}
  0\leq j_{L}\leq\frac{N}{2}\text{ \ \ and \ \ }0\leq j_{R}\leq\frac{N}{2} \,.
\end{equation}
In the target space interpretation, high angular momentum objects cease to
behave as point particles, and instead start puffing up to a large size
comparable to the IR cutoff.

We reach physical states not eliminated by the GSO\ projection by passing to
the descendants. This is obtained by acting with the lowering operators of the
left- and right-moving current algebra, $J_{-n}^{a}$, $\widetilde{J}_{-m}^{a}%
$, in the obvious notation. For example, the dilaton, metric and NS two-form
are captured by excitations of the form:%
\begin{align}
\left\vert \text{WZW\ Dilaton}\right\rangle  &  =J_{-1}^{a}\widetilde{J}%
_{-1}^{a}\left\vert j,m_L;j,m_R\right\rangle \, ,\label{dilatonWZW}\\
\left\vert \text{WZW Metric}\right\rangle  &  =J_{-1}^{(a}\widetilde{J}%
_{-1}^{b)}\left\vert j,m_L;j,m_R\right\rangle \, ,\\
\left\vert \text{WZW B-Field}\right\rangle  &  =J_{-1}^{[a}\widetilde{J}%
_{-1}^{b]}\left\vert j,m_L;j,m_R\right\rangle \, .
\end{align}
The \textquotedblleft dilaton\textquotedblright\ in particular can also be
viewed as the scale factor of an FRW model. In the
above, we have also imposed the level matching constraint which correlates the
angular momentum representation of the left- and right-movers.

The above description helps to explain the sense in which the F-theory model
UV\ completes our 3D M-theory description. The main point is that because we
have a solvable worldsheet CFT\ on the $S^{3}$ factor, we can track
the behavior of various states which appear in the M-theory limit. In
particular, we note that since the level $N=pq$ WZW\ model admits two
equivalent presentations as the $SU(2)/%
\mathbb{Z}
_{p}$ and $SU(2)/%
\mathbb{Z}
_{q}$ CFTs \cite{Gaberdiel:1995mx, Maldacena:2001ky}, we can take the extreme case where
$p=1$ and $q=N$. Said differently, we can work with either a fluxed $S^3$ or a fluxed $S^3 / \mathbb{Z}_N$

In the large $N$ limit the geometry of the untwisted sector is well-approximated by replacing
the $\mathbb{Z}_{N}$ factor by a continuous $U(1)$ factor. The resulting coset space $S^2 = SU(2)/ U(1)$ amounts
to collapsing the circle bundle of the Hopf fibration. From a physical perspective, we can thus work in terms of the $S^2$
geometry provided we also include all the twisted sector states as well. The fact that we ``lost a dimension'' motivates our proposal
to treat the projection onto the untwisted sector as the M-theory description.
This also shows that the M-theory limit is, by
itself, not UV\ complete, for the same reason that projecting onto the
untwisted sector of a CFT leads to a loss of unitarity in the full model.

\section{Supersymmetry in the Presence of Flux}\label{sec:KILLING}

In the previous section we presented a general proposal for how to make sense of F-theory
on $Spin(7)$ backgrounds. Our aim in this section will be to study the sense in which
supersymmetry can be preserved for such backgrounds. The main idea we develop here is
that although in signature $(3,1)$ this is not possible, in signature $(2,2)$ it is possible to have
``$\mathcal{N} = 1/2$ supersymmetry.'' Of course, in the analytic continuation back to Lorentzian signature
this means that any finite energy excitation will break supersymmetry. The only state which can enjoy
this ``$\mathcal{N} = 1/2$ supersymmetry'' is the ground state, which is the privileged state invariant
under all symmetry generators. For some recent discussion on the interpretation of
F-theory in $(10,2)$ signature, see reference \cite{Heckman:2017uxe}.

As we also need to make contact with various M-theory limits of this
construction, it will also prove helpful to study fluxed backgrounds for 3D vacua
generated from M-theory on $Spin(7)$ backgrounds. With this in mind, we first consider
the Killing spinor equations for M-theory backgrounds, and then turn to the
related analysis for F-theory backgrounds.

\subsection{M-theory on $\mathbb{R}_{\text{time}} \times S^2 \times Y_{8}$}

11D supergravity has a single 32 component Majorana spinor as a supersymmetry
parameter, $\epsilon$. On spacetimes which are warped products of the form:
\begin{equation}
\mathbb{R}_{\text{time}} \times S^2 \times Y_{8}
\end{equation}
the Lorentz group factors as a product and the supersymmetry parameter decomposes as
\begin{equation}
  \begin{aligned}
    Spin(10,1) &\rightarrow Spin(2,1) \times Spin(8) \rightarrow Spin(2)
    \times Spin(8) \cr
    {\bf 32} &\rightarrow ({\bf 2, 8^s}) \oplus ({\bf 2, 8^{s^{\prime}}}) \rightarrow
    {\bf 8^s}_{1} \oplus {\bf 8^s}_{-1} \oplus {\bf 8^{s^{\prime}}}_{1} \oplus {\bf
    8^{s^{\prime}}}_{-1} \,.
  \end{aligned}
\end{equation}
The ${\bf 2}$ representation of $Spin(2,1)$ is real, and the spinor
representations of the $Spin(8)$ can be taken to be individually Majorana due
to triality. When the holonomy of the internal manifold is taken to be reduced
to $Spin(7)$ then one finds
\begin{equation}
  \begin{aligned}
    Spin(8) &\rightarrow Spin(7) \cr
    {\bf 8^s} &\rightarrow {\bf 1} \oplus {\bf 7} \cr
    {\bf 8^{s^{\prime}}} &\rightarrow {\bf 8} \,,
  \end{aligned}
\end{equation}
and thus one can see that there will be exactly two real supercharges
preserved in this background.

The Killing spinor equation from the variation of the gravitino in 11D
supergravity is \cite{Cremmer:1978km}
\begin{equation}\label{eqn:MKS}
  \nabla_M \epsilon - \frac{1}{288}\left( \Gamma_M^{P_1\cdots P_4} - 8
  \delta_M^{P_1}\Gamma^{P_2\cdots P_4}\right)G_{P_1\cdots P_4}\epsilon = 0 \,,
\end{equation}
where $G$ is the M-theory four-form flux.
In the internal directions, along $Y_8$, we must be able to solve the Killing
spinor equations in such a way that $Y_8$ is a $Spin(7)$ manifold. As we have
shown that there is a single covariantly constant spinor, that is, one that
transforms trivially under parallel transport along $Y_{8}$, we need only to
solve the external Killing spinor equations, to guarantee that there is a
supersymmetric solution despite the non-trivial curvature of the $S^2$.

We decompose the 11D spinor in terms of components as
\begin{equation}
  \epsilon = (\rho_1^+ + \rho_1^-) \otimes \eta^s + (\rho_2^+ + \rho_2^-)
  \otimes \eta^{s^{\prime}} \,,
\end{equation}
where $\rho_i^\pm$ are the spinors transforming in the $Spin(2)$ spin
representation, and $\eta^{s,s^{\prime}}$ are the Majorana--Weyl spinors on the eight-manifold,
$Y_{8}$. Since $\eta^{s^{\prime}}$ does not give rise to a Killing spinor on $Y_{8}$ we
shall simplify the notation and write $\rho_1^\pm = \rho^\pm$. More generally,
we shall use $\rho$, sans superscript, to refer to the packaging of $\rho^\pm$
together into the ${\bf 2}$ of the broken $Spin(1,2)$.

We write the 11D metric as
\begin{equation}
  ds^2 = e^{2C} \left( - dt^2 + d\Omega_{S^2}^2 + ds^2_{Y} \right)
  \,,
\end{equation}
and decompose the 11D $\Gamma$ matrices in terms of the 3D and 8D $\Gamma$
matrices via
\begin{equation}
  \Gamma_\mu = \gamma_\mu \otimes \widetilde{\gamma}_9 \,, \quad \Gamma_m =
1_2 \otimes  \widetilde{\gamma}_m \,.
\end{equation}
We chose the four-form flux, $G_4$, such that it has one component which has
two legs along the putative sphere and two in the internal directions, and the
remaining contributions are supported entirely on the $Spin(7)$ manifold. We
write, schematically,
\begin{equation}
  G_4 = \epsilon_{\mu\nu}G_{mn} + G_{mnpq} \,.
\end{equation}
The external part of the Killing spinor equations then becomes
\begin{equation}
  \begin{aligned}
  \nabla_\mu \left(\rho_1^+ + \rho_1^-\right) - \frac{1}{288}\left( \Gamma_\mu^{P_1\cdots P_4} - 8
  \delta_\mu^{P_1}\Gamma^{P_2\cdots P_4}\right)G_{P_1\cdots P_4}\left(\rho_1^+ +
  \rho_1^-\right) = 0 \,.
  \end{aligned}
\end{equation}
On the 3D spacetime, the Killing spinor equation is simply:
\begin{equation}
(\nabla_{\mu} + \gamma^{\nu}F_{\nu \mu}) \rho = 0
\end{equation}
in the obvious notation. The upshot of this analysis is that at least in each local frame, there is
no issue with preserving two real supercharges on such fluxed backgrounds. Introducing
real supercharges $Q_{a}$ with $a = 1,2$, we have in a local frame:
\begin{equation}
  \{Q_{a} , Q_{b} \} = 2 \gamma^{\mu}_{ab} P_{\mu},
\end{equation}
with $P_{\mu}$ the local generators of translations, and
$\gamma^\mu$ are 3D gamma matrices.
The appearance of a warped product in $\mathbb{R}_{\text{time}} \times S^2 \times Y_{8}$
ensures that these supercharges are independent of time, that is, $[P_0 , Q_a] = 0$.

Globally, however, there is an important subtlety with preserving
supersymmetry with a round $S^2$ factor.  Indeed, since the isometry group is
$SO(3)$, we need our supercharges to assemble into pseudo-real representations
of the bigger symmetry algebra. This would lead to a contradiction with the
analysis just presented because the pseudo-real (rather than real)
representation would require four real degrees of freedom, not two.  So,
either we give up the isometries or the supersymmetries of the system.  It
seems clear that we need to give up the supersymmetries for all finite energy
excitations. However, there is no reason this also needs to hold for the
ground state. Indeed, if we consider the analytic continuation of our
spacetime to a different signature, we can define a supersymmetric theory with
two real supercharges and a ground state which \textit{is} annihilated by two
real supercharges. This ground state is the analog of the Bunch--Davies vacuum
in cosmology (see reference \cite{Bunch:1978yq}).  So, while all finite energy
excitations will have broken supersymmetry, the ground state can still enjoy
this feature.  Note also that this is really an infrared effect since frame by
frame, we can solve the Killing spinor equation.  So, locally we can still organize
fields according to 3D $\mathcal{N} = 1$
supermultiplets even though supersymmetry is broken by ``infrared effects'' in the spacetime.

With this caveat in mind, unitary representations of this algebra include a
3D $\mathcal{N} =1 $ vector multiplet
consisting of a 3D vector boson and a two component
real spinor. Additionally, we have the real
scalar multiplet consisting of a real scalar and
a two component real spinor.

\subsection{F-theory on $\mathbb{R}_\text{time} \times S^3 \times Y_{8}$}

We now turn to a similar analysis in the F-theory model. To frame our discussion,
recall that in the standard approach to Calabi--Yau
compactification in perturbative string theory, one instead seeks out geometries of the form $\mathbb{R}%
^{3,1}\times \text{CY}_3$. In this language, one specifies a covariantly constant
spinor on the $\text{CY}_3$ directions to retain unbroken supersymmetry in the
macroscopic spacetime dimensions. This is generalized in the context of
F-theory compactification, where the six internal dimensions are replaced by $B_{6}$,
the base of an elliptically fibered Calabi--Yau space.

To properly analyze all reality conditions for supersymmetric vacua
which descend from this higher-dimensional perspective, it is often
helpful to work in terms of the $32$-dimensional Majorana--Weyl spinor of
$Spin(10,2)$ and track its decomposition under the subalgebra $Spin(3,1)\times
Spin(7,1)$:
\begin{align}
Spin(10,2)  &  \supset Spin(3,1)\times Spin(7,1)\\
{\bf 32}  &  \rightarrow({\bf 2_{L},8^{s}})+({\bf 2_{R},8^{s^{\prime}}}) \,.
\end{align}
In this presentation, the first and second factors are complex conjugates. Indeed, the ${\bf 2_{L}}$
here labels a complex Weyl spinor.

From the perspective of F-theory, however, we typically want to view the $Spin(7)$ manifold as
a Riemannian manifold. That necessitates a different decomposition according to (by abuse of notation we use the same
labels for representations even though they have different reality properties):
\begin{align}\label{102spinoragain}
Spin(10,2)  &  \supset Spin(2,2)\times Spin(8)\\
{\bf 32}  &  \rightarrow({\bf 2_{L},8^{s}})+({\bf 2_{R},8^{s^{\prime}}}) \,,
\end{align}
where we then need to analytically continue in the $(2,2)$ signature directions.

Now, the reality properties of our spinors depend quite significantly on the signature of the spacetime.
For example, in signature $(3,1)$ the minimal amount of supersymmetry is a Weyl spinor, a complex doublet. In signature $(2,2)$,
however, we can simultaneously impose a Majorana and Weyl condition, so we can speak of a system with two real supercharges. In this sense,
it is valid to discuss $\mathcal{N} = 1/2$ supersymmetry. Note also that in the decomposition of line (\ref{102spinoragain}),
the eight-dimensional representations are real. In particular, we can independently decompose the $8^s$ and $8^{s^{\prime}}$.

So, at least in a 4D spacetime with signature $(2,2)$, we should not have any issue with preserving two real supercharges.
The physical interpretation will require us to analytically continue this back to signature $(3,1)$. The ``Bunch--Davies
vacuum'' of the system is still expected to be supersymmetric, since by definition it is required to be invariant
under all symmetries regardless of spacetime signature. Finite energy excitations are another matter altogether,
and this cannot be read off from an analytic continuation. For some additional discussion on this point, see for
example reference \cite{Polyakov:2007mm}. We also expect that in reduction to three dimensions we ought to
recover in each local frame 3D $\mathcal{N} = 1$ supersymmetry, so in this sense it is appropriate to speak of two real
supercharges.

To see how this comes about, we now trace through the Killing spinor analysis directly in signature $(3,1)$.
We find that in this signature, all supersymmetry is lost due to a reality condition on a pseudo-real spinor
representation. However, in the analytic continuation to $(2,2)$ signature, no such issue arises. We
also explain the sense in which each local frame in three dimensions retains two real supercharges.

In the spirit of F-theory, we geometrize possible seven-brane sources by
working in terms of an auxiliary 12D geometry.
Next, we apply the standard philosophy: We \textit{assume} that the
$Spin(7)$ background provides us with a solution to the internal Killing spinor equations.
Though obviously more difficult to solve, one can also formally state the requisite conditions
in terms of the IIB dilatino and gravitino variations (see Appendix \ref{app:IIB}).

Doing so, we find first of all, that the 4D fluxed Killing spinor equation
breaks into two conditions. First, we need to solve the Killing spinor equation in the presence of a torsion background:
\begin{equation}
(\partial_{\mu} + \omega^{(L)}_{\mu}(H)) \rho = 0 \, .
\end{equation}
Here, $\rho$ is a two-component left-handed Weyl spinor, and $\omega^{(L)}(H)$
is the left-handed spin connection in the presence of $H$-flux.
In the case of the WZW model, we have a specific choice of $H$-flux which allows us to immediately solve this equation.
Indeed, from our discussion near equation (\ref{leftshift}), the main effect is to
parallelize the local frame. We also note that the presence of a warped
product ensures that our Killing spinors are time independent.\footnote{A simple example of this type is
the near horizon geometry of an NS5-brane (see e.g. \cite{Callan:1991at}).}
In Appendix \ref{app:IIB} we consider a complexified three-form flux
as suggested by Type IIB supergravity. Here, we absorb this into an overall
choice of complex phase to maintain contact with our discussion
of WZW models presented earlier.

In addition to the 4D Killing spinor equation, the remnants of the $Spin(7)$ ansatz produce a projection
which relates the left-handed Weyl spinor $\rho_{\alpha}$ and its right-handed
conjugate $\rho^{\dag}_{\dot{\alpha}}$.
This is only possible because the presence of $H$-flux produces a new two
spinor index object $H^{\dot{\beta} \alpha}$:
\begin{equation}\label{projecto}
\rho^{\alpha} = \rho^{\dag}_{\dot{\beta}} H^{\dot{\beta} \alpha}.
\end{equation}
Absorbing all contributions from flux and warp factors into our definition
of this quantity, we can take it to be, up to a complex phase, the $2 \times 2$ identity matrix, that is to say:
\begin{equation}
H^{\dot{\beta} \alpha} = e^{i \theta}\overline{\sigma}^{\dot{\beta} \alpha}_{0}.
\end{equation}

Let us now state the supersymmetry algebra. In a local frame,
we introduce a two-component Weyl spinor $Q_{\alpha}$ and its conjugate
$Q^{\dag}_{\dot{\beta}}$. We then have the conditions:
\begin{align}
\{Q_{\alpha} , Q^{\dag}_{\dot{\beta}} \} & = 2 \sigma^{\mu}_{\alpha \dot{\beta}} P_{\mu} \\
Q^{\alpha} & = e^{i \theta} Q^{\dag}_{\dot{\beta}}
\overline{\sigma}^{\dot{\beta} \alpha}_{0} \,, \label{Qreality}
\end{align}
where $P_{\mu}$ is the local generator of translations. Additionally, our supercharges are independent of
time, that is, $[P_0, Q_\alpha] = 0$. It is sometimes also convenient to work
in terms of rigid supersymmetry. In that case, one should replace the spatial translations by the $SU(2)_{L}$ generators
of the $SU(2)_{L} \times SU(2)_{R}$ symmetries of the round $S^3$ (see references \cite{Festuccia:2011ws, Dumitrescu:2012ha}).
Let us also note that this analysis (aside from the reality condition of line (\ref{Qreality})) is in accord with that of
\cite{Festuccia:2011ws, Klare:2012gn}, namely a supersymmetric quantum field
theory can be placed on a three-sphere (with $H$-flux)\ without breaking any supersymmetry.
Additionally, we exploited the presence of the warp factor to ensure that our supercharges are independent
of time. Finally, the phase $e^{i \theta}$ specifies a choice of ``$\mathcal{N} = 1/2$ supersymmetry.''

How many supercharges do we now have? In fact, based on our analysis of the associated representations, we
can already see a potential issue with having $\mathcal{N} = 1/2$ supersymmetry in signature $(3,1)$.\footnote{We thank
T. Rudelius for communication on this point (as communicated to him by E. Witten).} Because the isometries of a round
$S^3$ ought to act on the supercharges, we should expect to get a representation of $Spin(4)$. But this is problematic because
the corresponding spinors transform in pseudo-real representations, so we would eventually wind up with four real degrees of freedom,
not two!

In fact, we expect that in this spacetime signature, supersymmetry will indeed appear to be broken, even in a local frame.
Returning to line (\ref{Qreality}), we see that
precisely because we are using the identity matrix to identify a $Q$ with its complex conjugate, we have
actually forced the $Q$'s to vanish. So in this sense,
the supersymmetry algebra in signature $(3,1)$ is totally vacuous.

But this is not so in signature $(2,2)$: In that case we can have Majorana--Weyl spinors with two real
degrees of freedom. The price we pay is that we have an unphysical spacetime signature for
any propagating (read finite energy) excitation. Nevertheless, we can use this analytic continuation
to define the ground state, and then proceed back to $(3,1)$ signature.
Indeed, all we demand is that our ground state be annihilated by the appropriate (complexified) symmetries.
Note that this is analogous to the definition of the Bunch--Davies vacuum via passing from Euclidean to Lorentzian signature \cite{Bunch:1978yq}.
So, while all finite energy excitations above the vacuum
will experience broken supersymmetry, the ground state will not. An additional comment is that in the associated WZW model, this involves
continuation from $SL(2,\mathbb{R})$ to $SU(2)$.

We can also now see from a 4D perspective how 3D $\mathcal{N} = 1$
supersymmetry fits into the story. To see why, consider the M-theory limit as obtained
by projecting onto the untwisted sector of the orbifold $S^3 / \mathbb{Z}_N$ in the large $N$ limit.
In a local frame, this reduction will look like we have chosen a preferential vector
in the spatial $\mathbb{R}^3$ factor. Returning to line (\ref{Qreality}), this introduces a further shift in the reality condition,
which we summarize as the replacement $\overline{\sigma}_{0} \rightarrow \overline{\sigma}_{3}$.
This in turn makes it possible to impose a non-trivial reality condition on the spinor, bringing us to 3D $\mathcal{N} = 1$ supersymmetry.

So, even though supersymmetry is broken, we expect the
field content of our 4D ``$\mathcal{N} = 1/2$'' theory to share many similarities with that of a 3D $\mathcal{N} = 1$ theory.
Viewing our $S^3$ as an $S^1$ bundle over $S^2$, we can first build multiplets on the base $S^2$ and then extend them into the fiber.
We defer further discussion of this to section \ref{sec:FTHEORY}.

\section{3D M-theory on a Local \boldmath{$Spin(7)$} Manifold \label{sec:MTHEORY}}

Though our eventual aim is the analysis of F-theory on warped products
$\mathbb{R}_{\text{time}} \times S^3 \times Spin(7)$ backgrounds, we first need to develop the
theory of the untwisted sector associated with the 3D M-theory model. Here, we study the case of M-theory compactified
on a local four-manifold of ADE singularities, and take the 3D spacetime to be $\mathbb{R}^{2,1}$.
We observe that each of these singular fibers can be presented in terms of a
non-compact elliptically fibered K3 surface, so we automatically get a torus fibration over a non-compact six-manifold.
The main complication is that this fibration is necessarily non-holomorphic to accommodate
the global structure of a $Spin(7)$ manifolds. This fibration gives rise
to a 7D super-Yang--Mills theory coupled to various spacetime filling defects.

So, our starting point will be a 3D $\mathcal{N}=1$ theory generated by M-theory compactified
on a local $Spin(7)$ manifold. For specificity, we assume that this $Spin(7)$
manifold takes the form of a non-holomorphic singular K3-fibration over a four-manifold $X_{\text{GUT}}$.
This can be arranged by taking a local $Spin(7)$ manifold given by an $\mathbb{R}^4$ bundle over $X_{\text{GUT}}$ and
then performing a quotient of each $\mathbb{R}^4$ fiber by $\Gamma_\text{ADE} \subset SU(2)$ a finite subgroup.

In particular, we do not require $X_\text{GUT}$ to
be a K\"{a}hler surface, but will require the existence of an orientation and
$Spin^{c}$ structure which is compatible with it being a calibrated cycle of
the Cayley four-form $\Omega_{(4)}$ of the $Spin(7)$ manifold. We allow for
further degenerations in the local K3 fibration to accommodate the analog of
matter and Yukawa couplings much as in the case of intersecting 7-branes obtained from Calabi--Yau fourfolds.

Now, from the perspective of M-theory on such a four-manifold, we see 7D
super-Yang--Mills theory on the space $\mathbb{R}^{2,1}\times X_{\text{GUT}}$.
The effective 3D theory equally has an interpretation in Type IIA string
theory as the theory on 6-branes wrapping a coassociative four-cycle,
$X_\text{GUT}$ inside of a (local) $G_2$ manifold.
To understand the low energy effective field theory in the 3D\ spacetime, we
follow the standard procedure of topologically twisting the theory on the
four-manifold.  To this end, we first consider the theory in flat space, and
after listing all of the fields and their symmetry transformation properties,
we apply the corresponding twist.

In flat space, 7D\ SYM\ has an $SU(2)_{I}$ R-symmetry, so this field
theory enjoys the global symmetries $SO(2,1)\times SU(2)_{L}\times
SU(2)_{R}\times SU(2)_{I}$. The field content under these different groups is:%
\begin{align}
\text{Flat Space Symmetries}  &  \text{: }SO(2,1)\times SU(2)_{L}\times
SU(2)_{R}\times SU(2)_{I}\\
\text{7D Gauge Field}  &  \text{:\ }({\bf 3,1,1,1})\oplus({\bf 1,2,2,1})\\
\text{Triplet of Scalars}  &  \text{: }({\bf 1,1,1,3})\\
\text{Gauginos}  &  \text{: }({\bf 2,2,1,2})\oplus({\bf 2,1,2,2})\text{.}%
\end{align}
Applying the Donaldson--Witten twist \cite{Witten:1988ze}, we embed the R-symmetry
$SU(2)_{I}\ $into the internal $SO(4)$ symmetry. Doing so, we get the
representation content of the twisted theory:%
\begin{align}
\text{Twisted Symmetries}  &  \text{: }SO(2,1)\times SU(2)_{L}\times
SU(2)_{\text{diag}}\\
\text{7D Gauge Field}  &  \rightarrow({\bf 3,1,1})\oplus({\bf 1,2,2})\\
\text{Triplet of Scalars}  &  \rightarrow\text{ }({\bf 1,1,3})\\
\text{Gauginos}  &  \rightarrow({\bf 2,2,2})\oplus({\bf 2,1,3})\oplus({\bf
2,1,1}) \,,
\end{align}
so as advertised, we produce a theory with a single doublet of supercharges on
the internal space, namely a 3D $\mathcal{N}=1$ supersymmetric theory. The
fields now assemble into supermultiplets of differential forms on $X_{\text{GUT}}$.
We denote these as a vector multiplet $V$ transforming as a 0-form on
$X_{\text{GUT}}$, a scalar multiplet $\mathbb{A}_{(s)}$ transforming as a 1-form
connection on $X_{\text{GUT}}$, and a scalar multiplet $\Phi_{(st)}$ transforming as
a self dual 2-form on $X_{\text{GUT}}$:
\begin{align}
  \text{ }V  &  \text{: }({\bf 3,1,1})\oplus({\bf 2,1,1})\\
  \mathbb{A}_{(s)}  &  \text{: }({\bf 1,2,2})\oplus({\bf 2,2,2})\\
  \Phi_{(st)}  &  \text{: }({\bf 1,1,3})\oplus({\bf 2,1,3}) \,.
\end{align}
We note that all bosons and fermions are counted via real degrees of freedom. Our convention for self-dual versus
anti-self-dual for the two-form scalars is dictated by the expected match to the special case where we take $X_{\text{GUT}}$
to be a K\"ahler surface. In that case, we ought to be able to pass to the Vafa--Witten twist, where there is an
adjoint valued $(2,0)$-form.

The bosonic action of 7D super-Yang--Mills theory is obtained by dimensional reduction of 10D super-Yang--Mills theory. Recall the
Lagrangian of that system is, for gauge group $G$:
\begin{equation}
L_\text{10D} = \frac{1}{g^2} \text{Tr} \left(\frac{1}{2} F \wedge \star F  -
\overline{\Psi} \, i \Gamma^{\mu}D_{\mu} \Psi \right) \,,
\end{equation}
with $F_{IJ}$ the 10D field strength and $\Psi$ a 10D Majorana--Weyl spinor in the adjoint representation.
This action is supersymmetric (on-shell) with the transformations (see e.g. \cite{Green:1987mn}):
\begin{align}
\delta_\text{10D SUSY} \Psi & = \frac{1}{2} \Gamma^{IJ} F_{IJ} \, \varepsilon \\
\delta_\text{10D SUSY} A_{I} & = i \, \overline{\varepsilon} \, \Gamma_{I}
\Psi \,.
\end{align}
Anticipating the expected contributions from supersymmetry, we write the 7D action in a particularly suggestive way using the
self-dual field strength $F_\text{SD}$ and the self-dual two-form $\phi_\text{SD}$ on $X_{\text{GUT}}$.
Assembling all terms according to 3D fields, we
can write the purely bosonic terms as a sum of kinetic and potential terms:
\begin{equation}
\mathcal{L}_\text{7D}^\text{bosonic} =
\mathcal{L}^\text{bosonic}_\text{kinetic} +
\mathcal{L}^\text{bosonic}_\text{potential} \,,
\end{equation}
with:
\begin{align}
\mathcal{L}^\text{bosonic}_\text{kinetic} & = - \frac{1}{2 g_\text{3D}^2} \text{Tr} \left( F_\text{3D} \wedge \star F_\text{3D} \right) - \vert D_\text{3D} A_\text{GUT} \vert^2
- \vert D_\text{3D} \phi_\text{SD} \vert^2 \, , \\
\mathcal{L}^\text{bosonic}_\text{potential} & = \int_{X_{\text{GUT}}} \, \left( - \vert F_\text{SD} + \phi_{\text{SD}} \times \phi_{\text{SD}} \vert^2
- \vert D_A \phi_\text{\text{SD}} \vert^2  \right) \, .
\end{align}
In the above, we have introduced a cross-product operation for self-dual two-forms. 
We automatically get this structure in the local model because the bundle of 
self-dual two-forms over $X_{\text{GUT}}$ is a non-compact $G_2$ manifold. 
As such, it comes equipped with an associative three-form $C_{(3)}$. 
Consequently, there is a natural pairing:
\begin{equation}
{\bigwedge}^{2}_\text{SD}(\text{ad}(P)) \times {\bigwedge}^{2}_\text{SD}(\text{ad}(P)) 
\rightarrow {\bigwedge}^{2}_\text{SD} (\text{ad}(P))
\,\,\,\text{with:}\,\,\,(\phi_{\text{SD}} \times \phi^{\prime}_{\text{SD}})_{k} =
C_{ijk} \phi^{i} \phi^{\prime j} \,,
\end{equation}
where here, we view the $\phi^{i}$ and $\phi^{\prime j}$ as triplets 
in ${\bigwedge}^{2}_\text{SD} \otimes \text{ad}(P)$, with $P$ a principal $G$-bundle. 
Note that $\phi_{\text{SD}} \times \phi_{\text{SD}}$ 
does not vanish because of the $\text{ad}(P)$ factor.

The terms involving the fermions of the action also follow directly from dimensional reduction of the 10D super-Yang--Mills theory
action. Our normalization conventions are fixed by matching to standard 3D $\mathcal{N} = 1$ and 4D $\mathcal{N} = 1$ supersymmetric actions.
We split this up into a kinetic term and interaction terms:
\begin{equation}
\mathcal{L}^\text{fermionic} = \mathcal{L}^\text{fermionic}_{\text{kinetic}} +
\mathcal{L}^{\text{fermionic}}_{\text{int}} \,.
\end{equation}
The kinetic terms implicitly depend on the metric of the internal geometry (which descends to the K\"ahler metric of the 3D $\mathcal{N} = 1$ theory), but are otherwise of the standard form. The interaction terms are dictated by the differential form content on $X_{\text{GUT}}$:
\begin{equation}
\mathcal{L}^\text{fermionic}_{\text{int}} = -\frac{1}{2}\chi_\text{SD} \wedge
D_{A} \psi_{(1)} + \frac{i}{\sqrt{2}}\star \psi_{(1)} \wedge D_{A} \eta_{(0)}
-\frac{1}{2} \psi_{(1)} \wedge [\phi_\text{SD} , \psi_{(1)}] +
\frac{i}{\sqrt{2}} \eta_{(0)} \wedge [\phi_\text{SD} , \chi_\text{SD}] +
\text{c.c.} \,.
\end{equation}

Indeed, the full 3D $\mathcal{N} = 1$ action can also be derived using the structure of the supersymmetry variations for the fields.
The on shell variations of the various fields are fully determined just by the condition that we build appropriate differential forms on $X_{\text{GUT}}$. Since the 10D super-Yang--Mills theory supersymmetry variations of the fermions always produce a 10D field strength, we only allow quadratic order terms for $\phi_\text{SD}$ and linear order terms in $F_\text{SD}$. Because we have a single real two-component doublet of supercharges in three dimensions, we can explicitly display the real spinor index. This essentially follows from Appendix D of reference \cite{Beasley:2008dc} as well as reference \cite{Apruzzi:2016iac} (suitably adapted) so we shall be brief:
\begin{align}
\delta_{a} \chi^\text{SD}_{b} & = i \sqrt{2} \gamma^{\mu}_{ab} D_{\mu} \phi_\text{SD}
- \sqrt{2} \epsilon_{ab} (F_\text{SD} + \phi_{\text{SD}} \times \phi_{\text{SD}}) \, , \label{deltachi} \\
\delta_{a} \psi^{(1)}_{b} & = i \sqrt{2} \gamma^{\mu}_{ab} D_{\mu} A^\text{GUT}_{(1)} - \sqrt{2} \epsilon \star \, D_{A} \phi_\text{SD} \, ,\\
\delta_{a} \eta^{(0)}_{b} & = \gamma^{\mu \nu}_{ab} F_{\mu \nu} \, .\label{deltaeta}
\end{align}
We obtain a 3D $\mathcal{N} = 1$ supersymmetric vacuum (classically) by considering background
values of the fields independent of the 3D spacetime, and also setting the remaining variations associated
with fluctuations on the four-manifold to zero.

In addition to these local interactions, one can in principle also include various Chern-Simons terms for the gauge fields, and this has been studied for example in reference \cite{Gukov:2001hf}. For an abelian
vector multiplet, this follows from reduction of the 11D topological terms $C_3 \wedge G_4 \wedge G_4$ and $C_3 \wedge I_8$.
For a non-abelian vector multiplet there is a natural extension of this which is fixed by the condition that it reproduces the abelian answer
in suitable limits. In the local setting the levels of the Chern-Simons theory are adjustable parameters. These terms play an important role in the quantum theory but we defer an analysis of this to future work.

Much as in other contexts which have been studied in the literature, including
\cite{Marcus:1983wb, ArkaniHamed:2001tb, Beasley:2008dc, Apruzzi:2016iac},
we can write down a supersymmetric action, integrated over the internal directions. Then, the
supersymmetric vacuum conditions reproduce the supersymmetric equations of
motion of our 7D gauge theory wrapped on $X_{\text{GUT}}$. This is by now a fairly
standard exercise in the F-theory / M-theory literature, so we
instead focus on the elements which are qualitatively different
compared with those cases.

The main thing we emphasize here is that there is a superpotential:
\begin{equation}\label{WMGUT}
W_{X}=\underset{X_{\text{GUT}}}{\int}\text{Tr}\left(  \Phi_{\text{SD}} \wedge
\left(\mathbb{F} + \frac{1}{3}\Phi_{\text{SD}} \times \Phi_{\text{SD}} \right)
\right)  \,,
\end{equation}
where:
\begin{equation}
\mathbb{F}=d\mathbb{A}+\mathbb{A}\wedge\mathbb{A}%
\end{equation}
is the field strength in the internal directions, presented as a 3D $\mathcal{N} = 1$ superfield.
Varying with respect to $\Phi_{\text{SD}}$ and $\mathbb{A}$, we
obtain the F-term equations of motion for the bosonic components:
\begin{equation}
D_{A}\phi_{\text{SD}} = 0\text{ \ \ and \ \ } F_\text{SD} + \phi_{\text{SD}}
\times \phi_{\text{SD}} = 0 \,,
\end{equation}
where in the second equation we used the fact that $\phi_{\text{SD}}$ is a
self-dual two-form.

Some of this analysis can also be recovered from a \textquotedblleft bulk
perspective.\textquotedblright\ Recall that a $Spin(7)$ manifold comes with a
distinguished real four-form $\Omega_{(4)}$, and that to retain supersymmetry,
we require our four-manifold $X_{\text{GUT}}$ to be a calibrated cycle with respect
to this four-form. Since we are also free to wrap M5-branes over other
four-cycles, we clearly get domain walls in the 3D $\mathcal{N}=1$ theory,
with a BPS\ tension set by the analog the Gukov--Vafa--Witten superpotential
\cite{Gukov:1999ya} (see also \cite{Acharya:2002vs}):%
\begin{equation}\label{Wbulk}
  W_{\text{bulk}}=\underset{Y_{8}}{\int}\Omega_{(4)}\wedge G_{(4)} \,,
\end{equation}
where $G_{(4)}$ denotes the difference in $G$-flux on the two sides of the
domain walls. For earlier work on string compactification and domain walls in
3D $\mathcal{N}=1$ theories, see for example in \cite{Witten:1999ds, Acharya:2001gy,Braun:2018fdp}.

We now see that at least the abelian subsector (that is, the part which can be seen purely geometrically)
of the superpotential of line \eqref{WMGUT} is in accord with
these considerations: Introducing harmonic two-form representatives
for the collapsing cycles of the ADE singularity, we write:
\begin{equation}
\delta \Omega_{(4)} \sim \phi_\text{SD} \wedge \omega_\text{fiber} \, , \qquad \delta G_{(4)} \sim F_\text{SD} \wedge \omega_\text{fiber} \, .
\end{equation}

\subsection{Bulk Zero Modes}

The physics of the 3D effective field theory involves the zero modes
generated by the theory on the four-manifold. These are obtained by first solving the F- and D-term constraints,
and then expanding around such a background. This breaks the original gauge symmetry
$G$ on the 6-brane down to some commutant subgroup $H \subset G$. At the level of the principal $G$-bundle $\text{ad}(P)$, this
involves a decomposition as:
\begin{equation}
\text{ad}(P )\rightarrow \bigoplus_{i} (\tau_{i} , \mathcal{E}_i) \, .
\end{equation}
where $\tau_{i}$ are representations of the unbroken group $H$.
The 3D $\mathcal{N} = 1$ zero mode content will then be determined
by the bundle content on the four-manifold. From the fluctuations
of the gauge field, we get zero modes in representations:
\begin{align}
\delta \mathbb{A}^{\tau_i} & \in H^1(X_{\text{GUT}} , \mathcal{E}_i) \, ,\\
\delta \mathbb{A}^{\tau_{i}^{\ast}} & \in H^1(X_{\text{GUT}} , \mathcal{E}^{\ast}_i).
\end{align}
For the fluctuations of the self-dual two-form we instead find:
\begin{align}
\delta \Phi^{\tau_i} & \in H^0(X_{\text{GUT}} , {\bigwedge}^{2}_\text{SD} \otimes \mathcal{E}_i) \, ,\\
\delta \Phi^{\tau_i^{\ast}} & \in H^0(X_{\text{GUT}} , {\bigwedge}^{2}_\text{SD} \otimes \mathcal{E}^{\ast}_i).
\end{align}
Observe that in the 3D $\mathcal{N} = 1$ theory, CPT conjugation amounts to switching self-dual and anti-self-duality conditions on the internal modes. This is reflected as a complex conjugation / dualization on the associated bundle assignments.

\subsection{Localized Matter and Interactions}

The study of localized matter and Yukawa interactions in our 3D theory also follows a
quite similar analysis. Much as in the case of F-theory GUTs, we can model these localized matter
fields as vortex solutions generated by a background value for the self-dual two-forms of the twisted 7D action. The main
thing we need to track is the equations for the trapped modes of the bulk theory. In flat space, we would identify these with
5D $\mathcal{N} = 1$ hypermultiplets. This again fits with the fact that in a perturbative IIA description
the geometric interpretation would be that of two 6-branes intersecting along a common curve.

Because of the restrictive nature of the differential form content,
the ``flavor symmetries'' of the system, in tandem with the condition that we go
only to quadratic order in the fermions dictates the resulting equations for trapped 5D modes:
\begin{align}
D_{A} \star \psi_{(1)} & = [\phi_\text{SD} , \chi_\text{SD}] \, ,\\
D_{A} \, \chi_\text{SD} & = [\phi_\text{SD} , \psi_{(1)}].
\end{align}
For example, in a local patch $\mathbb{R}^4 \sim \mathbb{C}^2$ the analysis is basically identical to that
presented in reference \cite{Beasley:2008dc}, but instead of 6D hypermultiplets wrapped on a curve we have 5D hypermultiplets.

To analyze the contribution of these localized matter fields to the 3D effective field theory, we
now show how to incorporate localized matter on curves in
this construction. We begin with a 5D hypermultiplet which fills
$\mathbb{R}^{2,1}\times\Sigma_{\text{matt}}$ where $\Sigma_{\text{matt}%
}\subset X_{\text{GUT}}$ is a Riemann surface. The matter field is in a generalized
bifundamental (in the usual sense of F-theory GUTs), and we suppress this in
what follows. Since the scalars of the hypermultiplet transform as a doublet
under the $SU(2)_{I}$ R-symmetry and the scalars are neutral, we learn that
after the twist, we have two superfields in conjugate representations
which transform as doublets in $K_{\Sigma}^{1/2}$, a spinor bundle on
$\Sigma_{\text{matt}}$. We note that because the matter curve is a Riemann surface, we can always
treat it as a complex curve, and count our modes in this way. We also note that much as in other contexts,
it may be necessary to switch on a background flux from both the bulk and flavor branes to cancel all anomalies.

Labelling these superfields as $\Lambda^{c}$
and $\Lambda$, we can also identify a coupling with the pullback of the bulk
gauge field onto the matter curve, as well as the pullback of the
self-dual two-form:
\begin{equation}\label{Wsigma}
W_{\Sigma}=\underset{\Sigma_{\text{matt}}}{\int}\Lambda
^{c}(d+\mathbb{A}_{\Sigma})\Lambda + (\Lambda^{\dag} \cdot \Phi_{\Sigma} \cdot
\Lambda - \Lambda^{c} \cdot \Phi_{\Sigma} \cdot \Lambda^{c \dag}) \,.
\end{equation}

The contributions from matter fields modifies the conditions to have a 3D supersymmetric
vacuum. Displaying the corrections from just one matter curve (they are summed in the full system of equations),
the new supersymmetric vacuum conditions are now:
\begin{align}
    D_{A} \phi_\text{SD} + \delta_{\Sigma} \, (\Lambda^c , \Lambda) & = 0  \, , \\
	F_\text{SD} + \phi_{\text{SD}} \times \phi_{\text{SD}} +  \delta_{\Sigma} \wedge \,
(\langle \Lambda^{\dag} , \Lambda \rangle - \langle \Lambda^{c} , \Lambda^{c \dag} \rangle) & = 0 \,.
\end{align}
where we have introduced canonical pairings $\left(\cdot , \cdot \right)$ and $\langle \cdot , \cdot \rangle $ associated with matter fields
valued in bundles $K_{\Sigma}^{1/2} \otimes \mathcal{U}$ and its dual:
\begin{align}
\left(\cdot , \cdot \right) & :  (K^{1/2}_{\Sigma} \otimes \mathcal{U}^{\ast}) \times (K^{1/2}_{\Sigma} \otimes \mathcal{U}) \rightarrow {\bigwedge}^{1}_{\Sigma} \, ,\\
\langle \cdot , \cdot \rangle & : (K^{1/2}_{\Sigma} \otimes \mathcal{U}^{\ast}) \times (K^{1/2}_{\Sigma} \otimes \mathcal{U}) \rightarrow {\bigwedge}^{0}_{\Sigma} \, .
\end{align}
They build up a triplet of moment maps for the 5D $\mathcal{N} = 1$ hypermultiplet. In the curved space considered here,
we view $(\cdot , \cdot)$ as a 1-form localized on $\Sigma_\text{matt}$ and $\langle \cdot , \cdot \rangle$ as a zero-form which
canonically pairs with the volume form of the curve. Additionally, variation of the matter fields
trapped on curves produces the zero mode equations in the background of the (pullback) of the bulk gauge field and self-dual two-form
to the curve:
\begin{equation}
	D_{A|{_\Sigma}} \Lambda + \phi|_{\Sigma} \cdot \Lambda^{c \dag} = 0 \,\,\,\text{and}\,\,\,
D_{A|{_\Sigma}} (\Lambda^c)^\dag + \phi|_{\Sigma} \cdot \Lambda^{\dag}= 0.
\end{equation}
When $\phi_{\text{SD}}$ vanishes on the curve, the zero mode counting for
localized modes is therefore identical to what one has for standard
F-theory GUTs \cite{Beasley:2008dc}.

Yukawa interactions between matter fields localized on curves can be obtained by starting from the superpotential
for bulk modes on the four-manifold, and then expanding to cubic order in fluctuations about the background. This yields
terms localized at points $p$ of the internal four-manifold of the schematic form:
\begin{equation}
W_\text{Yuk} = \delta_{p} \,\, \Lambda^{(1)} \cdot \Lambda^{(2)} \cdot \Lambda^{(3)}.
\end{equation}

\subsection{Spectral Cover}

\begin{figure}[t!]
\begin{center}
\includegraphics[trim={0cm 2cm 0cm 0cm},clip,scale=0.45]{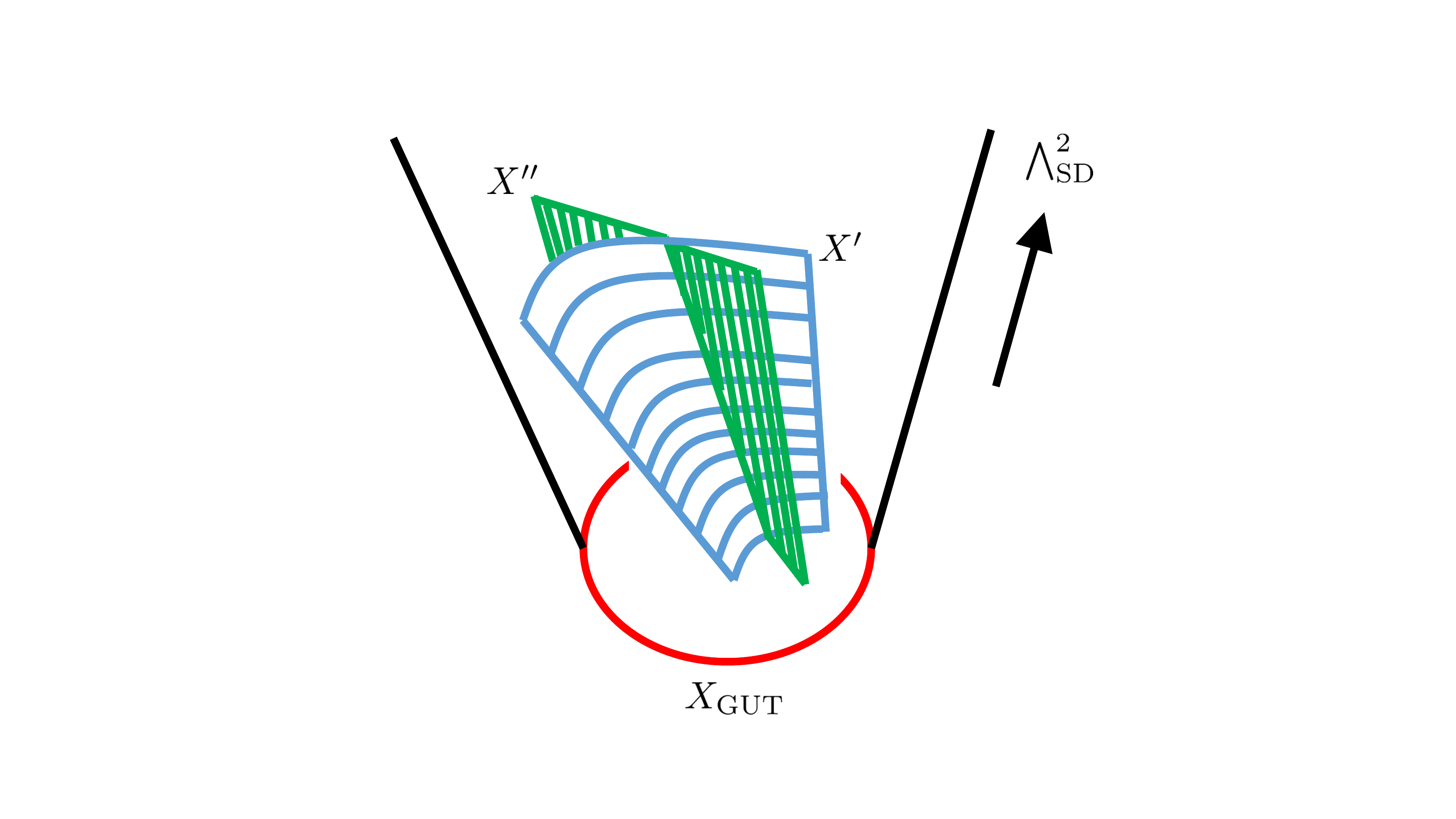}
\end{center}
\caption{Depiction of the spectral cover description obtained from 7D super-Yang--Mills
theory wrapped over a four-manifold $X_{\text{GUT}}$, as governed by the
Donaldson--Witten twist. The total space is a non-compact $G_2$ manifold given by the
bundle of self-dual two-forms over $X_{\text{GUT}}$. Pairwise intersections
lead to matter localized on Riemann surfaces and triple intersections
produce Yukawa couplings.}
\label{fig:SpectralCover}
\end{figure}

Gauge theory on the $6$-brane automatically
generates a spectral cover construction. Here, our Higgs fields
are in the bundle of self-dual two-forms ${\bigwedge\nolimits_\text{SD}^{2}}
\rightarrow X_{\text{GUT}}$ which specifies a local $G_{2}$ manifold! In this
characterization, we have local coordinates $u^i$ for the normal
directions and build the corresponding spectral cover by a polynomial in
$u^i$ with coefficients given by the Casimir invariants of $\Phi$. For
example, in the case of a 7D gauge theory with gauge group $SU(N)$, the
spectral equation is:%
\begin{equation}
\det(u^i-\phi^i)=0
\end{equation}
in the obvious notation. This sweeps out a four-manifold in the local
$G_{2}$ space. See figure \ref{fig:SpectralCover}
for a depiction.

In matching to the bulk geometry where we expect a local model with a non-holomorphic K3 fibration over a four-manifold,
we can also see where this geometry fits into this picture. There is a canonical construction of the
twistor space $\text{Tw}(X_{\text{GUT}})$ of $X_{\text{GUT}}$ by
restricting to unit norm self-dual two-forms, namely a six-manifold. We note
that in the cases at hand, we typically need to delete some locus from the four-manifold to
produce the actual match from the local to bulk dynamics. This in particular means that the
twistor fibration comes with a non-holomorphic section. This $S^2$ fiber can then serve the role as the base of a
non-holomorphic K3 fibration.

\begin{figure}[t!]
\begin{center}
\includegraphics[trim={0cm 4cm 0cm 2cm},clip,scale=0.45]{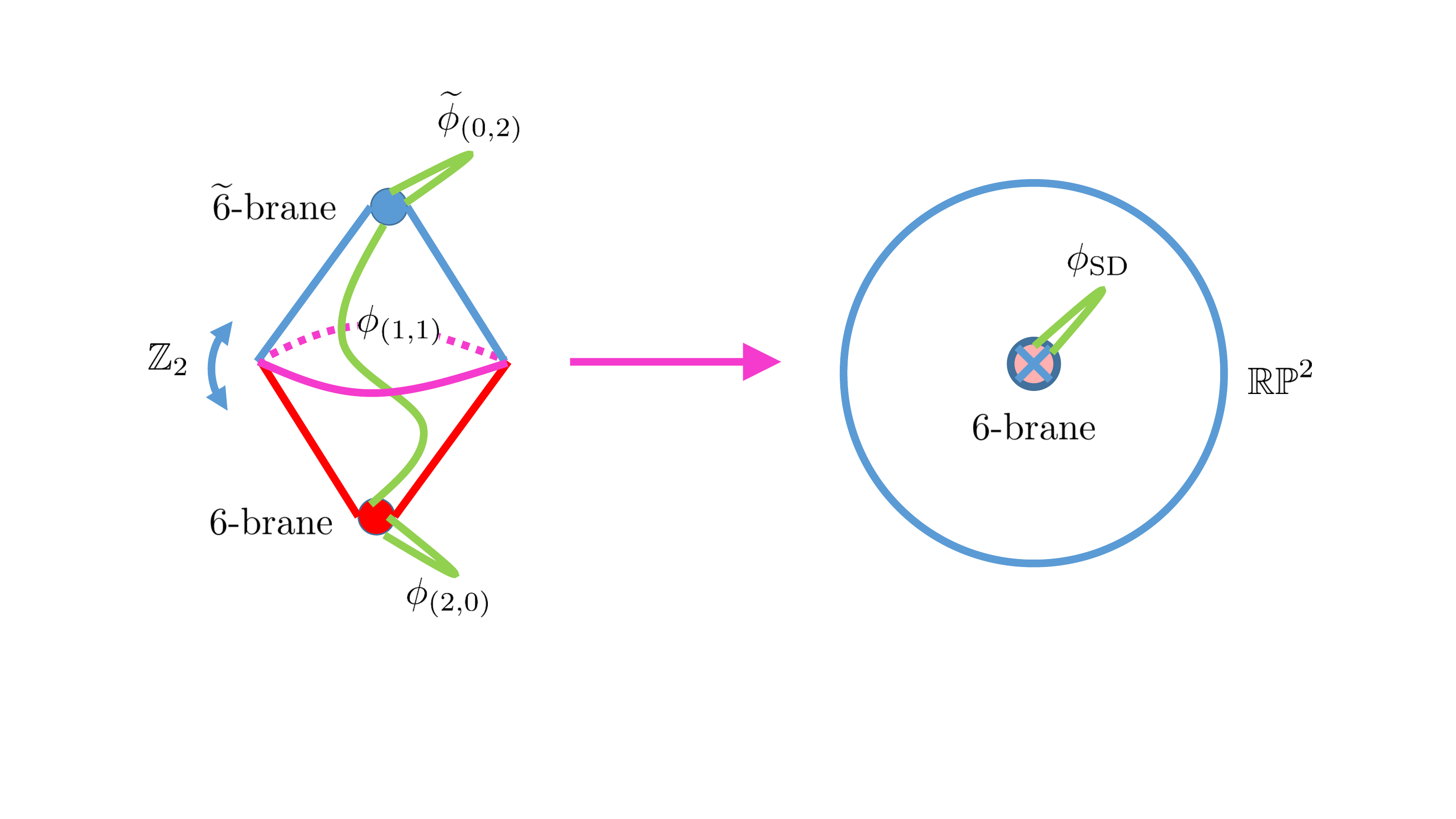}
\end{center}
\caption{Depiction of the local model generated by gauge theory on a $6$-brane wrapped on a four-manifold $X_{\text{GUT}} = S_\text{GUT}$ in the limit of 3D $\mathcal{N} = 2$ supersymmetry (left) and its generalization to 3D $\mathcal{N} = 1$ supersymmetry (right). In the case of enhanced supersymmetry, there is a local
Calabi--Yau fourfold, and the moduli of the $Spin(7)$ model correspond to a real slice through these parameters. In the corresponding spectral
cover construction, this real slice amounts to a $\mathbb{Z}_2$ identification of two spectral covers with a fixed $5$-cycle ``at infinity.'' In the limit where the $6$-brane sits on top of the $\mathbb{Z}_2$ invariant locus one obtains a different twisted gauge theory on the four-manifold. The local geometry experienced by a $6$-brane wrapped on such a four-manifold is an $\mathbb{RP}^2$ bundle over $X_{\text{GUT}}$.}
\label{fig:66braneimage}
\end{figure}

In the 7D\ gauge theory, however, there is a clear sense in which the internal
equations of motion are classically scale invariant. This
suggests a different compactification by identifying all points of the normal
coordinate $u^i$ up to real rescalings, namely an $\mathbb{RP}^{2}$ (an $S^2$ with
an orientation reversing crosscap inserted). See figure \ref{fig:66braneimage} for a depiction.

In either case, we see that in both the global picture and the local
picture, it is also natural to view our spectral four-manifold as moving in an
ambient six-manifold. In particular, this means that when we consider
intersections of $X_{\text{GUT}}$ in our six-manifold, the intersection takes place
over Riemann surfaces, and triple intersections take place over points in
$X_{\text{GUT}}$.

\subsection{Specialization to 3D $\mathcal{N} = 2$ Supersymmetry \label{sec:specialization}}

It is also instructive to track what happens in the special case where we assume additional supersymmetry with the local
geometry a Calabi--Yau fourfold and the four-manifold a K\"ahler surface $S_\text{GUT}$. Then, we should expect a 3D $\mathcal{N} = 2$ theory, with the internal gauge theory governed by the Vafa--Witten twist. The first observation is that the fields now assemble into $(p,q)$-differential
forms of the complex manifold. In particular, we can now work in terms of a $(0,1)$ connection for the gauge group, and the triplet of scalars decomposes to an adjoint valued $(2,0)$-form and a self-dual $(1,1)$ form on $S_\text{GUT}$. The interpretation of
the $(1,1)$ form requires particular care.

For K\"ahler manifolds the $(2,0)$ forms are self-dual and there is a distinguished self-dual $(1,1)$ form proportional to the
K\"ahler class which we write as $J_{S}$.
Setting the Coulomb branch scalar to zero, the supersymmetric vacuum conditions
(\ref{deltachi}) -- (\ref{deltaeta}) now imply:
\begin{align}
F_{(0,2)} & =0\label{deltachi2} \, ,\\
\overline{\partial}_{A} \phi_{(2,0)} & = 0 \, , \\
J_{S} \wedge F_{(1,1)} + \frac{i}{2} [\phi_{(2,0)} , \phi_{(0,2)}^{\dag}] & = 0 \, , \label{deltaeta2}
\end{align}
which conforms with the supersymmetry variations for 7-branes wrapped on a K\"ahler surface
as obtained in reference \cite{Beasley:2008dc}.

We can also see the geometric origin of the adjoint valued $(1,1)$-form. Returning to our discussion of the
superpotential interaction of line \eqref{Wbulk}, we observe that a Cayley four-form can be constructed from
the holomorphic four-form and the K\"ahler form via \eqref{eq:Cayley_fourform_from_CY4}:%
\begin{equation}\label{idento}
	\Omega_{Spin(7)}= \operatorname{Re}(e^{i\theta} \, \Omega_{\text{CY}_{4}}) + \frac{1}{2} J \wedge J \, .%
\end{equation}
The phase $e^{i\theta}$ specifies an $\mathcal{N}=1$ subalgebra of the
3D $\mathcal{N}=2$ theory.
Consider next local variations of $\Omega_{\text{CY}_{4}}%
$. In the fourfold, these are given by a $(3,1)$-form which we split into a
piece  $\phi_{(a,b)}$ on the K\"{a}hler surface $S_\text{GUT}$ and a piece in the fiber directions,
labelled by harmonic forms $\omega_{(p,q)}$:%
\begin{equation}
\delta\Omega_{(3,1)} \sim \phi_{(2,0)}\wedge\omega_{(1,1)}. \label{localvariation}%
\end{equation}
In a $Spin(7)$ manifold, this cannot be disentangled from variations of $J \wedge J$:
\begin{equation}\label{JJvariation}
\delta (J \wedge J) \sim \phi_{(1,1)} \wedge \omega_{(1,1)}.
\end{equation}
Indeed, the $\phi_{(2,0)}$, its conjugate $\phi^{\dag}_{(0,2)}$ along with
$\phi_{(1,1)}$ combine to form the triplet of anti-self-dual three-forms on
$X_{\text{GUT}}$.

We can also ask how to match the corresponding spectral cover constructions of the two models. In the 3D $\mathcal{N} = 2$ case,
the local geometry experienced by the 7D super-Yang--Mills theory is $\mathcal{O}(K_{S}) \rightarrow S_\text{GUT}$. As already remarked, the local spectral cover construction in the generic $\mathcal{N} = 1$ case is the bundle of self-dual two-forms over $S_\text{GUT}$. What we are doing
is performing a projection down along one of the rays of this 7D geometry to reach the local Calabi--Yau threefold. From this perspective, the branes in $\mathcal{O}(K_{S}) \rightarrow S_\text{GUT}$ appear to come in pairs, which are to be identified under a $\mathbb{Z}_2$ involution as dictated by the choice of reality condition
in line \eqref{idento}. To accomplish this projection whilst retaining the local Calabi--Yau threefold geometry we wrap
-- in perturbative Type IIB terminology -- an O5-plane on the five-cycle given by $S_\text{GUT} \times S^{1}$, where the $S_1$ is the circle
``at infinity'' in the normal direction of $\mathcal{O}(K_{S}) \rightarrow S_\text{GUT}$. The geometry is thus better viewed as an $\mathbb{RP}^2$ bundle over $X_{\text{GUT}}$, in accord with our earlier remarks on the spectral cover construction.

We label these as a $6$-brane and its image $\widetilde{6}$-brane. Note also that there are three sets of modes parameterizing the motion of the $6$-brane in the Calabi--Yau geometry, with form content dictated by the topological twist:
\begin{align}
\begin{split}
6 - 6 \text{ strings:} \quad & \text{adjoint valued (2,0)-form}, \\
\widetilde{6} - \widetilde{6} \text{ strings:} \quad & \text{adjoint valued (0,2)-form}, \\
6 - \widetilde{6} \text{ strings:} \quad & \text{adjoint valued (1,1)-form}.
\end{split}
\end{align}
While the first two sets of modes are massless in the 3D $\mathcal{N} = 2$ theory, the last set is massive and is integrated out. Observe, however, that in the special limit where the $6$-brane and $\widetilde{6}$-brane coincide, this additional mode becomes massless, in accord with
the mode content dictated by the 3D $\mathcal{N} = 1$ theory. See figure \ref{fig:66braneimage} for a depiction of this geometry.

\section{4D F-theory Lift \label{sec:FTHEORY}}

Let us now turn to the 4D physics associated with F-theory compactified on
this local $Spin(7)$ geometry. As already mentioned, the key link between the
M-theory and F-theory pictures is that we exploit the exact duality between
the level $N$ WZW\ models on $S^3 / \mathbb{Z}_N$ and $S^3$.
In particular, if we start with a 3D\ effective theory on
$\mathbb{R}_{\text{time}}\times S^{2}$, this should really be viewed as a
projection onto the untwisted sector of the full model. To capture this
additional data, we need to find a way to incorporate all of the
twisted sectors of the $S^3 / \mathbb{Z}_N$ model, or equivalently
its abelian T-dual description as an $S^3$.

At a pragmatic level we need to track the fate of our various 3D fields as we attempt to lift them to 4D fields.
To begin, consider a 3D vector field on the spacetime $\mathbb{R}_{\text{time}} \times S^2$. Clearly, we
extend it to depend on the $S^{1}$ Hopf fiber of the full $S^{3}$. Doing so,
all 3D fields are automatically promoted to a 4D vector field on $\mathbb{R}_{\text{time}} \times S^{3}$. So, we summarize this
by the schematic substitution:
\begin{equation}
A_\text{3D}(x_{S^{2}},x_{\text{GUT}})  \rightarrow  A_\text{4D}(x_{S^{2}},x_{S^{1}},x_{\text{GUT}}) \, . \label{A3D4D}
\end{equation}
Consider next the lift of the 3D $\mathcal{N} = 1$ vector multiplet. We have the original 3D gaugino, so the natural
guess is that this will turn into a 4D Weyl fermion. This indeed is what we ought to expect from acting with all the isometries of
the $S^3$ on our field. Perhaps more directly, we note that in equation (\ref{A3D4D}), we have picked up another component to the gauge field. Splitting these as $A_\text{4D} \sim A_\text{3D} \oplus A_{\bot}$, we can treat $A_{\bot}$ as the scalar component of its own 3D $\mathcal{N} = 1$ multiplet. This in turn means that we have identified another 3D spinor. These two spinors combine to produce a single 4D Weyl spinor.

Similar considerations hold for the real scalar multiplets. Indeed, observe that in the absence of Chern-Simons terms, we can dualize a scalar to a vector field. Then, the analysis just presented for vector fields applies and we pick up an additional set of partners. That is to say, the 4D lift will produce another real degree of freedom and another 3D spinor. Of course, from a 4D perspective, we are not free to dualize the original real scalar to a 4D vector. This in particular means that the 4D lift must be packaged in terms of a complex scalar, so we get the same degrees of freedom as a 4D $\mathcal{N} = 1$ chiral multiplet. We note that similar subtleties apply even in ``ordinary'' circle compactifications of F-theory to 3D M-theory models, as in reference \cite{Grimm:2010ks}.

The one caveat in this sort of analysis is that it could be that these lifts are not independent of one another. For example, in the case of a 4D $\mathcal{N} = 1$ theory reduced on a (flat) circle to a 3D $\mathcal{N} = 2$ theory, the Coulomb branch scalar of the vector multiplet can alternatively be viewed as its own 3D $\mathcal{N} = 1$ multiplet. These subtleties are simple to identify, however, because the only candidate scalars where this could occur must transform in the adjoint representation of the gauge group.

Putting this together, we see that in the analysis of the previous section, the configuration of intersecting 6-branes will automatically produce the field content expected from a configuration of intersecting 7-branes with $\mathcal{N} = 1$ supersymmetry. In particular, if we attempt to engineer the Standard Model on such a configuration, we should also expect all of the standard superpartners of the MSSM.

That being said, though the \textit{field content} is the same as an $\mathcal{N} = 1$ supersymmetric theory, the actual interactions will
definitely violate $\mathcal{N} = 1$ supersymmetry. This is quite apparent in the 3D $\mathcal{N} = 1$ theory where standard 4D flat space considerations such as holomorphy do not exist. It is thus better to view our theories as having ``$\mathcal{N} = 1 / 2$ supersymmetry.''\footnote{There is an unfortunate clash of terminology with that of reference \cite{Seiberg:2003yz} which introduced a notion of half supersymmetry upon making the Grassmann coordinates of superspace have non-trivial anti-commutators. The two notions are not related.} As
we have already remarked several times, supersymmetry is completely broken for all finite energy excitations.

From the perspective of F-theory compactification, however, this at first
presents a puzzle. Recall that in supersymmetric compactifications of
F-theory, we identify each component of the discriminant locus with a 7-brane gauge
theory, namely 8D super-Yang--Mills theory wrapped on divisors of the base and
filling our spacetime. In 8D super-Yang--Mills, the bosonic sector includes an
8D gauge boson and two real scalars. The above lift to include the twisted
sectors produces an 8D gauge boson, but also includes a \textit{triplet} of real
scalars, not a pair of scalars. Somehow we have wound up with one more scalar
than in 8D super-Yang--Mills!

The key point is that in situations with such low supersymmetry and such low
dimension, the intuition of more (super)symmetric situations can fail. Returning to
our discussion near line \eqref{JJvariation}, we already observed that there is
another mode from the variation of $J \wedge J$:
\begin{equation}
\delta(J \wedge J) = \phi_{(1,1)}\wedge\omega_{(1,1)}.
\end{equation}
This additional mode from $\phi_{(1,1)}$ cannot be decoupled from the other
localized modes of the 7-brane. Indeed, as we have already remarked
in section \ref{sec:specialization}, the non-holomorphic K3 fibration should really
be viewed as defining a configuration of 7-branes and its anti-holomorphic image. The
additional $7 - \widetilde{7}$ string is this $(1,1)$-form.

\subsection{Backreaction and $\mathcal{N} = 1 \rightarrow \mathcal{N} = 1/2$ Breaking}\label{sec:LOCAL}

The considerations in the previous sections already provide
strong evidence that there is no a priori obstruction to realizing F-theory
backgrounds with $Spin(7)$ manifolds. That being said, for the purposes of model building
one would like to make use of the extensive literature now available
in realizing particle physics models in F-theory (for early reviews
see references \cite{Heckman:2010bq, Weigand:2010wm}).

It is beyond the scope of the present work to build a complete phenomenology.
Indeed, we have already reviewed the state of the art for constructing
$Spin(7)$ manifolds in section
\ref{sec:REVIEW}. We anticipate that the spectral cover construction presented in section \ref{sec:MTHEORY}
will provide the requisite tools to start constructing and analyzing these geometries.

With this in mind, our aim in this subsection we will be somewhat more modest. We shall ask whether we can identify
the appropriate ingredients of 4D $\mathcal{N} = 1$ models and how they are modified in going from F-theory on a Calabi--Yau
fourfold to a $Spin(7)$ manifold. An additional byproduct of our
analysis is that it can clearly accommodate the particle physics sector of an
F-theory GUT\ model of the kind introduced in
\cite{Beasley:2008dc,Donagi:2008ca}.

To begin, we consider F-theory on an elliptically fibered Calabi--Yau fourfold
$Y\rightarrow B$ with $B$ a K\"{a}hler threefold base. This is described by a
Weierstrass model of the form:%
\begin{equation}
y^{2}=x^{3}+fx+g,
\end{equation}
with $f$ and $g$ respectively sections of $K_{B}^{-4}$ and $K_{B}^{-6}$. The
discriminant:%
\begin{equation}
\Delta=4f^{3}+27g^{2}%
\end{equation}
indicates much of the physical data of an F-theory model. A component of the
vanishing locus of the discriminant $\left(  \Delta=0\right)  $ indicates the
location of K\"{a}hler surfaces wrapped by 7-branes. Collisions of these
components indicate localized matter on complex curves and triple
intersections indicate the locations of Yukawa couplings.

In a local model, we assume that we have a GUT\ group localized on a
contractible K\"ahler surface $S_{\text{GUT}}$. The local geometry in the
neighborhood of this 7-brane takes the form of a non-compact Calabi--Yau
fourfold (really a log-Calabi--Yau space see e.g. \cite{Donagi:2012ts})
dictated by a $dP_{9}$ fibration over $S_{\text{GUT}}$:
\begin{equation}
	\begin{tikzcd}[row sep = normal, column sep= normal  ]
		dP_9 \arrow[hook, r] & \text{CY}_4^\text{local} \arrow[d] \\
		& S_\text{GUT}
	\end{tikzcd} \, .
\end{equation}
The $dP_{9}$ is itself an elliptically fibered surface, and degenerations in
the elliptic fibration produce the requisite gauge group, matter content and
Yukawa couplings in the local patch.

Now, if we treat both the fiber and base as compact, we discover that the
total space will not be Calabi--Yau. Rather, the candidate holomorphic
four-form will have a pole along a divisor. This divisor is an elliptically
fibered Calabi--Yau threefold CY$_{3}^{\text{Het}}$ with base $S_{\text{GUT}}$.
Deleting this threefold from the geometry, we reach a non-compact
log-Calabi--Yau space. The reason for the superscript is that in compact models which
admit a stable degeneration limit, this threefold is to be identified with a
dual heterotic Calabi--Yau threefold. Here, we shall remain agnostic as to the
existence of a global heterotic dual.

\begin{figure}[t!]
\begin{center}
\includegraphics[width = \textwidth]{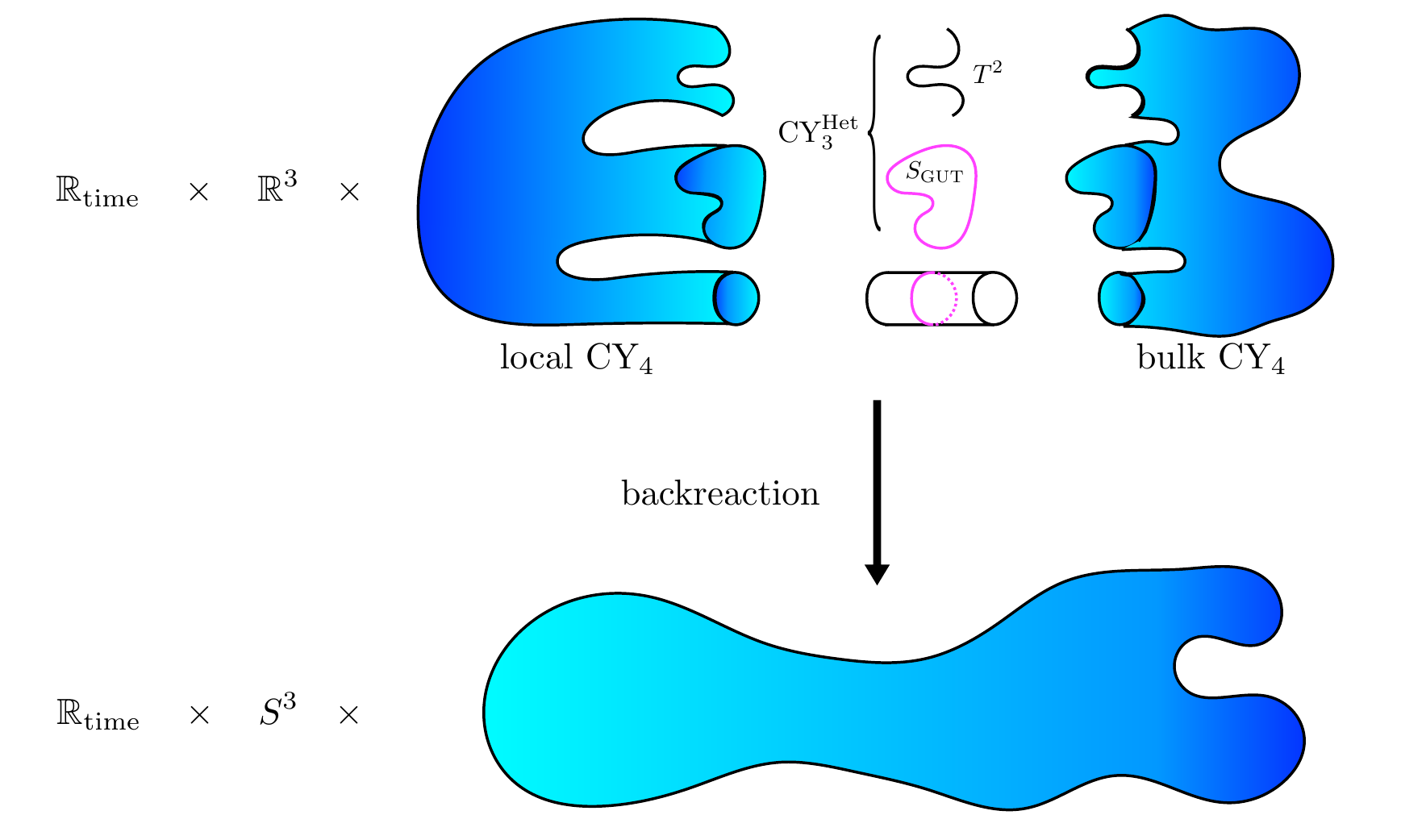}
\end{center}
\caption{Depiction of the proposed physical construction of $Spin(7)$ manifolds using elliptically fibered Calabi--Yau fourfold building blocks, glued along a $\text{CY}_3^\text{Het}$ in an asymptotically cylindrical region.
In the Type IIB picture, we wrap NS5-branes on a local 5-cycle $S_\text{GUT} \times S^1$, which is stable against perturbations because of a running dilaton profile.
After taking into account backreaction,
the NS5-branes dissolve into flux, leaving behind a $Spin(7)$ geometry with a fluxed $S^3$ in the 4D spacetime.}
\label{fig:backreact}
\end{figure}

The asymptotic behavior of the Calabi--Yau metric near this deleted region
takes the form of an asymptotic cylindrical region (see e.g. \cite{TianYauI, TianYauII, BandoKobayashi})
with this $\text{CY}_{3}^{\text{Het}}$ fibered over a cylinder:%
\begin{equation}\label{cylindrical}
\begin{tikzcd}
	\text{CY}_3^\text{Het} \arrow[d] \\
	S^1 \times \mathbb{R}_\perp
\end{tikzcd} \, ,
\end{equation}
in the obvious notation. See figure \ref{fig:backreact} for a depiction of the local geometry
in this cylindrical region. A very important feature of this
local construction is that the only components of the discriminant locus which
appear in this cylindrical region are associated with $I_{1}$ fibers which are
associated with the presence of perturbative D7-branes. In particular, no
non-perturbative branes \textquotedblleft leak out\textquotedblright\ into
this region. Denoting the local coordinate on the $\mathbb{R}_{\bot}$ factor
by $x_{\bot}$, the main idea is that in the region $x_{\bot}\rightarrow
-\infty$ we attach to the original particle physics sector, and in the region
$x_{\bot}\rightarrow+\infty$ we instead glue into the rest of a compact
fourfold. Doing so, we would produce a 4D $\mathcal{N}=1$ theory in flat
space.

We now ask what happens when we wrap NS5-branes of
Type IIB\ string theory over the 5-cycle $S_{\text{GUT}}\times S^{1}$ defined
by the cylindrical region of line \eqref{cylindrical}. For specificity, we
place them in the region $x_{\bot}\rightarrow\infty$ of the $\mathbb{R}_{\bot
}$ factor.

In the original 4D geometry, these branes sit at some marked point of
space, and fill the temporal direction $\mathbb{R}_{\text{time}}$. First of
all, we note that in this local geometry, there is indeed a supersymmetric 5-cycle
available to us, so the resulting configuration will break half of the
original supersymmetry, leaving us with two real supercharges. In the compact
model with base a K\"ahler threefold, we do not expect to generically have such
5-cycles. However, we ought to remember that the presence of these NS5-branes
also lead to a throatlike region where the dilaton will appear to become quite
large. This in particular means that the 5-brane is locally stable against
\textquotedblleft slipping off\textquotedblright\ into the bulk of the
Calabi--Yau fourfold. Turning the discussion around, we also see that because there are
no homologically non-trivial 5-cycles in the \textit{compact} base,
there is also no Gauss' law constraint on these 5-branes. This is similar
in spirit to the mismatch between local and global homology cycles
\cite{Buican:2006sn, Beasley:2008kw, Donagi:2008kj}.

In perturbative Type IIB terms, the appearance of a running dilaton is actually
crucial. Recall that in our discussion of the
WZW model near line \eqref{deficit}, we had a mild deficit in the central
charge $\delta c=6/(N+2)$, as well as a fluxed three-sphere. The appearance
of the NS5-branes is the most straightforward way to get a fluxed $S^3$ with a running dilaton.
Going to a slightly sub-critical dimension just means the NS5-brane picks up a small thickness.

With these caveats dealt with, let us now turn to the near horizon geometry in
the presence of our NS5-branes. The directions transverse to the NS5-brane
include the $\mathbb{R}_{\text{space}}^{3}\times\mathbb{R}_{\bot}$ directions.
Out of these, we see that there is a local throat region with an $S^{3}$
factor. One can view this as being generated by smearing the NS5-branes over
these directions. This smearing is really a shortcut for understanding the
effects of backreaction of the brane solution on the full geometry. Indeed,
the near horizon geometry of NS5-branes has already been worked out in
reference \cite{Callan:1991at} so we can simply adapt the relevant results to
the present case. We use notation as in the excellent review in the Appendix
of reference \cite{Schulz:2011ye}. After taking into account the effects of backreaction,
we expect a local geometry with a $\mathbb{R}_{\text{time}} \times S^3$ factor and
with local dilaton profile in the internal directions:
\begin{equation}
	\log \operatorname{Im}\tau_\text{IIB}\sim x_{\bot} / \sqrt{q},
\end{equation}
for $q$ a constant set by the worldsheet beta function for the WZW model.
Note that in these sorts of solutions, there are inevitably
regions where the dilaton will blow up. Here, we can appeal to an
S-dual frame. At a more pragmatic
level, we can also simply cut off the profile of the dilaton, and appeal to
the existence of a suitable smoothing of the full backreacted geometry. We
leave this point for additional analysis. Let us note that in the 10D analysis of the Killing
spinor equations presented in Appendix \ref{app:IIB}, we also need to assume a suitable profile for the dilaton
in the internal directions. Again, the simplest possibility is a contribution from the backreacted limit of 5-branes.

Finally, there is a non-zero $H$-flux threading the $S^{3}$ given by the
condition:%
\begin{equation}
\frac{1}{2\pi}{\int_{S^3}}H = 2\pi N\text{ \ \ and \ \ }%
d\star(\operatorname{Im}\tau_\text{IIB} \, \, H)=0.
\end{equation}
Here, we must assume that after backreaction, the original branes have
dissolved into flux and deform the original geometry. There is essentially
only one candidate available compatible with the appearance of an
$\mathbb{R}_{\text{time}}\times S^{3}$ factor with two real supercharges:\ The internal
directions of the candidate F-theory model must be a $Spin(7)$ manifold! It
would clearly be interesting to analyze the resulting asymptotic metrics,
perhaps along the lines of \cite{Cvetic:2001pga,Cvetic:2001ye,Cvetic:2001zx}.

Before closing this section, let us also comment on the relation of these $\mathcal{N} = 1$ building blocks
to those of the GCS construction of $Spin(7)$ manifolds given in \cite{Braun:2018joh}. In
reference \cite{Braun:2018joh}, the crucial ingredient of one example involves a further
specialization of CY$_{3}^{\text{Het}}$ to the Schoen manifold. This can be
viewed as a $T^{4}$ fibration over a $\mathbb{P}_{\text{GCS}}^{1}$. In the
GCS\ construction, one performs surgery using this $\mathbb{P}_{\text{GCS}%
}^{1}$ factor rather than the cylindrical region we have instead used. Indeed,
in the GCS\ construction, we have 7-branes wrapping the full $\mathbb{P}%
_{\text{GCS}}^{1}$ which would clearly clash with the presence of NS5-branes.
It would be interesting to see whether our starting
point can be used to produce additional GCS-like constructions of $Spin(7)$
manifolds.

\section{Closed String Sector}\label{sec:CLOSED}

In this section we study some aspects of the closed string sector. This
includes both the reduction of the higher-dimensional metric as well as all
$p$-forms and their superpartners. The gravitational sector for M-theory on
Calabi--Yau fivefolds has been considered in \cite{Haupt:2008nu}, and a related
analysis has also been performed in reference \cite{Lawrie:2016rqe}. For the most part,
this analysis goes through unchanged in the present case.
Additionally, determining the general
structure of the effective quantum mechanics will be complicated by
contributions from fluxes and instantons. On the flip side, because of
the low amount of supersymmetry, many moduli should automatically be
stabilized.

Putting aside these model dependent issues, there is a universal closed string
modulus which is certainly important to the evolution of the Universe and
figures prominently in the quantum evolution of our system:\ It is the scale
factor of the FRW Universe! Indeed, focussing on the 4D
spacetime, we have found a solution which is essentially the Einstein static
Universe but with an exotic form of stress energy. A curious feature of
the standard Einstein static Universe is that it is actually unstable against
small perturbations:\ The spacetime will classically either collapse to small
size or expand forever to large size. Both features are intriguing, and from a
phenomenological perspective this can indeed form the starting point for
various cosmological scenarios such as \cite{Ellis:2002we, Barrow:2003ni, Ellis:2003qz}.

To study some aspects of the resulting cosmology, we follow the general spirit of compactification
and dimensionally reduce our 4D\ gravitational system on the $S^{3}$ to a 1D
quantum mechanics problem. This is also the philosophy of the Wheeler--DeWitt equation \cite{DeWitt:1967yk,osti_4830161} and
its truncation in the mini-superspace approximation \cite{Hartle:1983ai}. We view
the spatial components of our 4D metric as dynamic quantum fields and use the
foliation by time to write down a corresponding quantum equation governing
these spatial components. Our plan in this section will be to track this
behavior in the reduction to one dimension.

With this in mind, our starting point is the 4D action:%
\begin{equation}
S_\text{grav, 4D}=\frac{1}{16\pi G_{N}}\int d^{4}x\,\sqrt{-g}\,\left[  R-\frac{1}%
{2}|H_{3}|^{2}-2\Lambda\right]  \,.
\end{equation}
Here we actually passed to Einstein frame as opposed to the discussion in section
\ref{sec:MFMF} and by abuse of notation we will keep the same notation for the
various fields as the one used in string frame.
The cosmological constant piece comes from the presence of a dilaton profile in
the internal dimension. We will reduce all terms separately. In all cases we
will need the form of the metric:
\begin{equation}
ds^{2}=-\mathcal N(t)^2\,dt^{2}+a(t)^{2}d\Omega_{(3)}^{2}\,.
\end{equation}
For convenience of the reader we recall that in our conventions the metric of a
unit radius $S^{3}$ is:%
\begin{equation}
d\Omega_{(3)}^{2}=d\psi^{2}+\sin^{2}\psi\,d\theta^{2}+\sin^{2}\psi\sin
^{2}\theta\,d\phi^{2}\,.
\end{equation}

Let us now turn to the dimensional reduction of the 4D\ action to one
dimension. To this end, we first compute the Ricci scalar for this
metric. The Einstein--Hilbert term is then:
\begin{equation}
\sqrt{-g}R=3\frac{a\sin\theta\sin^{2}\psi}{\mathcal N^{3}}\left[  2\mathcal N^{4}-2a\dot{a}%
\mathcal N\dot{\mathcal N}+2\mathcal N^2\left(  \dot{a}^{2}+a\ddot{a}\right)  \right]  \,.
\end{equation}
For simplicity we may integrate by parts so that we get rid of the $\ddot{a}$
term%
\begin{equation}
\frac{a^{2}\ddot{a}}{\mathcal N}=-2\frac{\dot{a}^{2}a}{\mathcal N}+%
\frac{\dot{\mathcal N}\dot{a}a^{2}}{\mathcal N^{2}}+\dots\,,
\end{equation}
where the ``$\cdots$'' includes a total derivative. Therefore the Einstein--Hilbert term of the
action becomes
\begin{equation}
S_{\text{EH}} = \frac{1}{16\pi G_{N}}\int d^{4}x\,\sqrt{-g}\,R = \frac{3\pi}{4G_{N}}\int dt\left[  a\mathcal N-\mathcal N^{-1}a\dot{a}^{2}\right]
\end{equation}

We need to follow a similar procedure for dimensional reduction of the
contributions from the $H$-flux and the effective cosmological constant from the
internal profile of the dilaton. Doing so, we get:%
\begin{align}
 \int_{S^3} H_3 \wedge \star_{S^3} H_3  &= \mathcal N \frac{N^2 l_s^4}{2 \pi^2 a^3} \,,\\
-\frac{1}{8\pi G_{N}}\int\sqrt{-g}\,\Lambda &  =-\frac{\pi}{4G_{N}}\int
dt\,a^{3}\mathcal N\,\Lambda.
\end{align}
Putting all of this together, we obtain the 1D\ action for the scale factor:%
\begin{equation}
S_\text{1D}[a]=\frac{3\pi}{4G_{N}}\int dt\left[  a\mathcal N-\frac{\mathcal N}{a^{3}%
}\frac{N^{2}l_{s}^{4}}{48\pi^{4}}-\mathcal N^{-1}a\dot{a}^{2}-\frac{1}{3}%
\mathcal N\,a^{3}\,\Lambda\right]  \,.\,
\end{equation}

\subsection{Classical Equations of Motion} \label{ssec:classico}

We can compute the equations of motion from this action. Varying with respect
to $\mathcal{N}(t)$ we obtain%
\begin{equation}
\frac{\dot{a}^{2}+1}{a^{2}}=\frac{N^{2}l_{s}^{4}}{48\pi^{4}a^{6}}+\frac{1}%
{3}\Lambda\,.\,
\end{equation}
Varying with respect to $a$ we find%
\begin{equation}
\frac{\ddot{a}}{a}+\frac{1}{2}\left(  \frac{\dot{a}}{a}\right)  ^{2}+\frac
{1}{2a^{2}}=\frac{\Lambda}{2}-\frac{N^{2}l_{s}^{4}}{32\pi^{4}a^{6}}\,.
\end{equation}
In both equations we eventually made the gauge choice $\mathcal N(t)=1$. These are the
usual Friedmann's equations with some exotic matter.\footnote{To recover the
usual presentation of Friedmann's equation one should take $1/2$ of the first
and subtract it from the second.} The matter we have has the following energy
density and pressure
\[
\rho=p=\frac{1}{128\pi^{5}G_{N}}\frac{N^{2}l_{s}^{4}}{a^{6}}\,.
\]
So this matter has $w=1$ consistent with the scaling $\rho\sim a^{-6}$. This is sometimes
referred to as a stiff fluid.

\begin{figure}[t!]
\begin{center}
\includegraphics[trim={0cm 2.5cm 0cm 3cm},clip,scale=0.5]{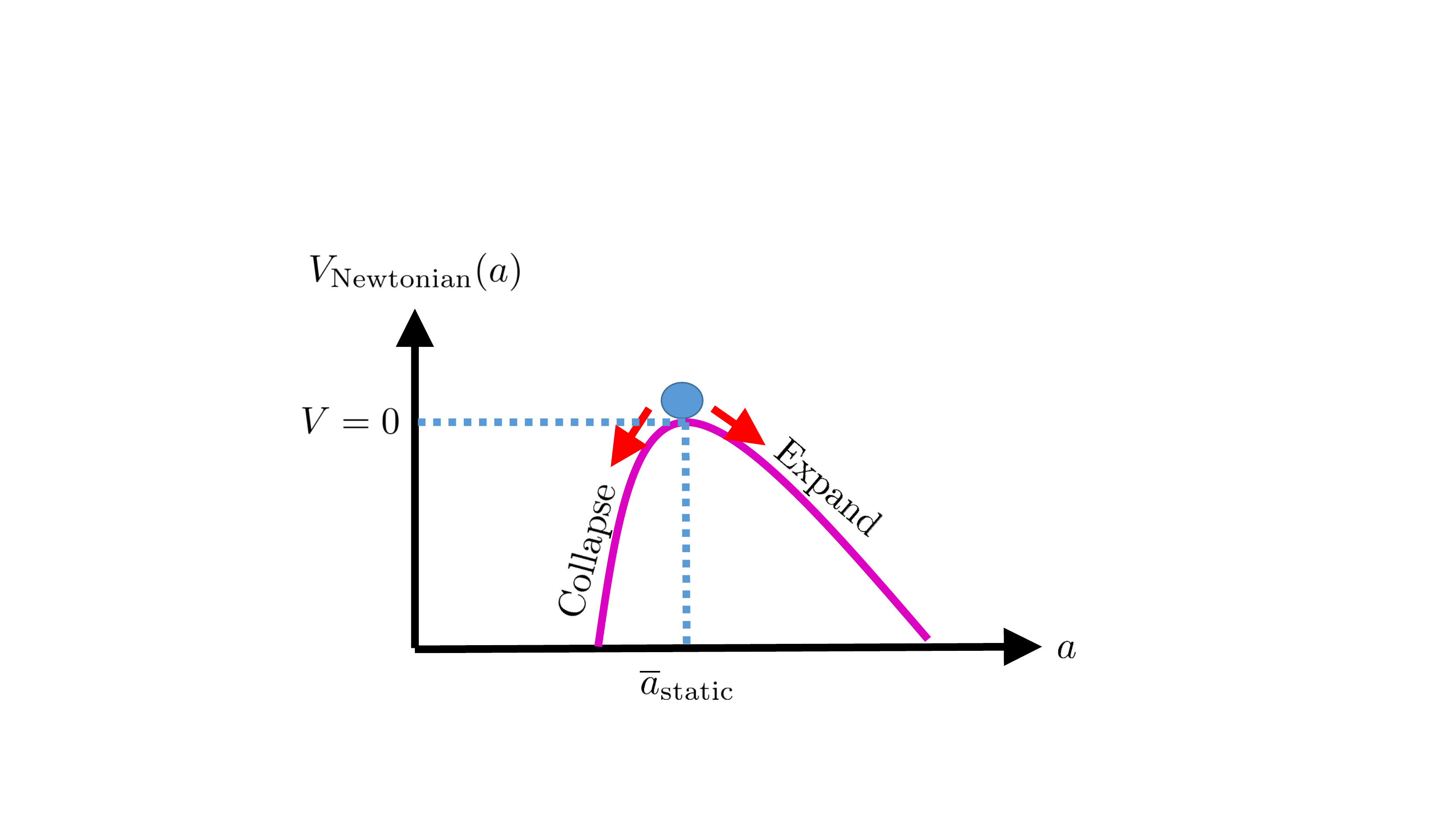}
\end{center}
\caption{Depiction of the instability of the effective Newtonian potential for the scale factor $a$ in our model. Much as in the case
of the Einstein static Universe, this can lead to either a collapsing or expanding Universe.}
\label{fig:RipCrunchNew}
\end{figure}

One may look for static solutions where $\dot{a}=0$. The solution is
\begin{align}
\Lambda &  =16\pi G_{N}\rho=\frac{8\pi^{2}}{l_{s}^{2}N}\,,\label{eq:cc}\\
\label{eq:abar}\bar{a}^{2}  &  =\frac{Nl_{s}^{2}}{4\pi^{2}}\,.
\end{align}
To check whether the solution is stable or not we can perturb it. We keep the
cosmological constant fixed to the value in \eqref{eq:cc}. Then in the
effective field theory we get a potential of the form
\begin{equation}\label{eq:pot}
V(a)=-a+\frac{N^{2}l_{s}^{4}}{48\pi^{4}a^{3}}+\frac{1}{3}a^{3}\frac{8\pi
^{2}}{l_{s}^{2}N}\,.
\end{equation}
The value \eqref{eq:abar} is indeed an extremal point of the potential and
at this point the vacuum energy vanishes:
\begin{equation}
V(\bar{a}) = 0,
\end{equation}
which is compatible with the fact that at the extremal point the
solution is supersymmetric. In addition to this we have that
\begin{equation}
V^{\prime\prime}(a)|_{a=\bar{a}}=\frac{16\pi}{l_{s}N^{1/2}}>0\,.
\end{equation}
One might be tempted to say that this system is stable around the vacuum, but
this is not the case. In fact the Lagrangian has the wrong sign in the
kinetic terms, and therefore when comparing with the kinetic energy it is
better to say that the correct Newtonian potential is minus the one appearing in
\eqref{eq:pot}. This leads to an unstable system which might either collapse
or expand (see figure \ref{fig:RipCrunchNew}). Indeed, the instability of the
Einstein static Universe has been appreciated for some time \cite{1930MNRAS}.

Based on our analysis of sections \ref{sec:MTHEORY} and \ref{sec:FTHEORY} we see no a priori obstruction
to also including matter and radiation in more realistic models.
Though we leave a discussion of fully fledged phenomenological scenarios for future work, we
observe that in the expanding phase, we automatically produce a model which will be dominated by dark energy.
This suggests a natural connection to some aspects of reference \cite{Agrawal:2018own}, though as far as we can tell, there is
no need for a quintessence field. Additionally, we see that if the system actually tips over into a collapsing phase,
the symmetry $a \rightarrow - a$ suggests the system will roll back out or tunnel quantum mechanically to
large scale factor again. So, another appealing feature of this sort
of cosmology is the relative insensitivity to initial conditions. As a final comment, we note that
even if the compactification moduli has other runaway directions, the volume of the $S^3$
decompactifies faster, so some issues with moduli stabilization would also be alleviated.

\subsection{Supersymmetry and Runaways}

One might ask whether the appearance of this instability in our spacetime might be ``cured'' by including supersymmetry. First of all,
the sense in which we have supersymmetry at all is extremely mild. Indeed, it only properly exists if we
analytically continue to signature $(2,2)$.

Even if the 1D quantum mechanics had an emergent $\mathcal{N} = 2$ supersymmetry, we still do not
expect it to lift such runaways. For example, in the related context of M-theory compactified on Calabi--Yau fivefolds,
the resulting moduli also exhibit runaway behavior \cite{Haupt:2008nu}. Indeed, a
common issue in many moduli stabilization scenarios
is the ``unwelcome'' feature that typically, not all moduli can be stabilized
and in fact runaway behavior leads to a decompactification limit. From
the considerations presented previously, we see that not only is
such runaway behavior to be expected, it is actually welcome!

With this in mind, let us briefly review the related case of decompactification instabilities
which already appears in the case of 1D supersymmetric quantum mechanics obtained
from M-theory on a Calabi-Yau fivefold. In this case, the volume moduli assemble into 2A supermultiplets, and as such,
are governed by a K\"ahler potential and a superpotential. For a brief review of 1D supersymmetric quantum mechanics,
see Appendix \ref{app:SQM}. The first question we need to address is whether it is compatible with supersymmetry to
have a ``wrong sign kinetic term'' for the scale factor. The sigma model metric for the volume moduli is not quite the moduli space metric on the fivefold, but is instead shifted by an outer product of the moduli. Labelling these moduli as $V_i$, the sigma model K\"ahler potential metric takes the schematic form:
\begin{equation}
G_{ij}^\text{sigma} = G_{ij}^\text{geometric} - c V_i V_j,
\end{equation}
where $c$ is an order one constant, and the
resulting metric has precisely one negative eigenvalue.
Explicit examples of runaway behavior, as generated by
this sort of metric are presented in \cite{Haupt:2008nu, Haupt:2009hw}.

In the case of M-theory compactified on the warped product of $S^2 \times Y_8$, we again anticipate the same indefinite signature form for
the sigma model metric. The moduli involve the volume of the $S^2$, as well as various two-cycles in $Y_8$. In the F-theory description, we instead have the moduli $\mathrm{Vol}(S^3)$ and $\mathrm{Vol}(B_6)$, the base of the torus fibered $Y_8$, the dilaton, as well as many other moduli. Of course, to really extract the actual physical action one ought to perform a systematic dimensional reduction of the model, perhaps
using a suitably adapted version of the analysis presented in reference \cite{Bonetti:2013fma}.

\subsection{Quantum Scale Factor}

Let us now turn to the quantum mechanics associated with this scale factor. We
recall that the 1D\ action is:%
\begin{equation}
S_\text{1D}[a]=\frac{3\pi}{4G_{N}}\int dt\left[  a\mathcal N-\frac{\mathcal N}{a^{3}%
}\frac{N^{2}l_{s}^{4}}{48\pi^{4}}-\mathcal N^{-1}a\dot{a}^{2}-\frac{1}{3}%
\mathcal N\,a^{3}\,\Lambda\right]  \,.\,
\end{equation}
In the full string compactification, we ought to view $1/G_{N}$ as
proportional to the volume modulus of the internal directions. This
contributes another dynamical mode but we shall assume that the dynamics of
this can be treated adiabatically.

The first perhaps surprising feature of this action is that the sign of the
kinetic term for the scale factor appears to be associated with a ghostlike
degree of freedom. This is a generic feature of the Wheeler--DeWitt equation
and occurs inevitably in trying to write a metric on the moduli space of
spatial metrics \cite{DeWitt:1967yk,osti_4830161}. The same issue also appears in compactification
of M-theory on Calabi--Yau fivefolds \cite{Haupt:2008nu}. This of course
leads to significant complications in the standard
field theoretic interpretation of the wave function of the Universe which is
an ongoing matter of some contention. For some discussion of the different
proposals taken for how to deal with this issue see for example
\cite{Gibbons:1978ac, Mazur:1989by, Halliwell:1989dy}.
Perhaps the simplest answer is that in the context
of 4D gravity, we can simply discard this mode as non-dynamical quantum
operator, as in \cite{Mazur:1989by}, namely it is an artifact of us dropping the other
contributions to the metric. In our case, this will not really work, because
we always must pay attention to the dilaton. More precisely, the relevant mode is, in the
language of the WZW\ model, captured by the zero mode on the $S^{3}$ as in
(\ref{dilatonWZW}), the dilaton mode of the WZW\ model at zero momentum:%
\begin{equation}
\left\vert a^{2}\right\rangle =\frac{1}{3} g_{S^3}^{bc} J^{-1}_{b}\widetilde{J}^{-1}_{c}\left\vert
0,0;0,0\right\rangle ,
\end{equation}
where $g_{S^3}^{bc}$ is the metric of a unit 3-sphere.
In particular, this means that the resulting quantum
theory will admit an excitation associated with this mode. In the Hilbert
space of the 2D worldsheet theory, this is a normalizable mode:\ There is no
sense in which it appears to be pathological at all.

What then, should we make of the apparently wrong sign kinetic term? In the
effective field theory approach, this suggests adopting a different
quantization scheme for the scale factor. This can be accomplished in the path
integral by the choice of a different contour of integration for this field
$a(t)$. Different choices lead to wildly different behaviors for the wavefunction
(see e.g. \cite{DiazDorronsoro:2017hti, Feldbrugge:2018gin}). We view the present discussion as a
promising starting point for addressing some of these issues.

\section{Witten's Dark Fantasy Revisited}\label{sec:FANTASY}

As already mentioned in the Introduction, one of the motivations for
developing the formalism of F-theory on $Spin(7)$ backgrounds is that it
provides a potential way to simultaneously resolve several 4D\ cosmological
issues. In particular, it could help to explain how \textquotedblleft%
$\mathcal{N}=1/2$ supersymmetry\textquotedblright\ might cancel off the zero
point energy of a quantum field theory, whilst still permitting a split mass
spectrum between bosons and fermions. The main bottleneck in implementing
Witten's proposal \cite{Witten:1994cga, Witten:1995rz, Becker:1995sp} and
Vafa's extension to F-theory \cite{Vafa:1996xn} is that it has been difficult
to make sense of the 4D physics. Here we argue that the construction of the
previous sections provides a framework for implementing \textquotedblleft
Witten's dark fantasy.\textquotedblright

The main idea we utilize is that the 4D F-theory vacuum on $\mathbb{R}%
_{\text{time}}\times S^{3}$ has a supersymmetric ground state annihilated by
two supercharges (in the analytic continuation to $(2,2)$ signature), and
similar statements hold for the 3D M-theory vacuum on $\mathbb{R}%
_{\text{time}}\times S^{2}$.

We expect that the mass of the superpartners will be controlled by some
expression involving $\ell_{\text{4D}}$ the 4D Planck length and $L$ the radius of
the $S^{3}$. Out of these quantities, we can also obtain an estimate for the
vacuum energy density in terms of the radius of the $S^{3}$ and the Planck
length:%
\begin{equation}
\rho_{\text{vac}}\sim\left(  \frac{1}{\ell_{\text{4D}}L}\right)  ^{2}%
\equiv\frac{1}{\ell_{\text{IR}}^{4}}\text{,}%
\end{equation}
where we have introduce a derived IR\ mass scale as specified by
$\rho_{\text{vac}}$. From the WZW\ model, there is also a
natural relation between $L$ and $\ell_{\text{4D}}$ given by:%
\begin{equation}
L^{2}\sim N\ell_{\text{4D}}^{2},\label{WZWradius}%
\end{equation}
and the T-dual Hopf fiber in $S^{3}/%
\mathbb{Z}
_{N}$ has average length $\widetilde{L}=L/N$, i.e.:%
\begin{equation}
L\widetilde{L}\sim\ell_{\text{4D}}^{2}.\label{Tduality}%
\end{equation}
In terms of this $N$ scaling, we also have:%
\begin{equation}
\ell_{\text{IR}}^{2}\sim\sqrt{N}\ell_{\text{4D}}^{2}.\label{IR}%
\end{equation}

Let us now turn to the expected mass splitting in three dimensions and its
extrapolation to four dimensions. Consider then, the 3D theory defined by
M-theory on the warped product $\mathbb{R}_{\text{time}}\times S^{2}\times
Y_{8}$. Though there is no issue with retaining supersymmetry in each local
frame, it is broken by an \textquotedblleft infrared effect.\textquotedblright We have already explained in section
\ref{ssec:classico} that at the critical point for the scale factor, the total
energy is zero in the quantum mechanics system. Additionally, since we still
have a supersymmetric ground state, the zero point energies of all quantum
fields will exactly cancel. Even so, there is a quite mild mass splitting in
the bosonic and fermionic degrees of freedom induced by gravitational effects.
As explained in \cite{Witten:1994cga, Witten:1995rz}, this is because any
finite energy excitation generates a conical deficit angle, obstructing the
existence of a globally conserved supercharge. The mass splitting then takes
the form:
\begin{equation}
\Delta m_{\text{3D}}\sim\frac{\kappa_{\text{3D}}^{2}}{L^{2}}%
,\label{deltamsplit}%
\end{equation}
with $\kappa_{\text{3D}}^{2}\sim\ell_{\text{3D}}$ the 3D\ Planck Length. Here,
the 3D Einstein--Hilbert action takes the form:
\begin{equation}
S_{\text{EH}}=\frac{1}{2\kappa_{\text{3D}}^{2}}\int\,d^{3}x\,\sqrt
{-g_{\text{3D}}}\,R_{\text{3D}}.
\end{equation}
Provided perturbation theory in $\kappa_{\text{3D}}^{2}$ is valid, we see that
there is a reliable approximation with an extremely small mass splitting
between bosonic and fermionic masses. Witten's dark fantasy is that there
exists a 4D\ \textquotedblleft strong coupling limit\textquotedblright\ in
which we continue to use the mass splitting formula of line
\eqref{deltamsplit}, as it is protected by 3D supersymmetry.

In fact, this is exactly what we expect to happen based on the way we got our
4D\ background in the first place! To see why, consider dimensionally
reducing the 4D theory along the Hopf fiber. The proper way to carry this is out
is to work in terms of the T-dual description so that we reduce on a small
circle of size $\widetilde{L}=L/N$. Additionally, we must allow for some
non-trivial $N$ scaling because we must account for the presence of many
twisted sector states. So, on general grounds, we have a relation of the
form:
\begin{equation}
\frac{1}{\ell_{\text{3D}}}\sim\frac{1}{N^{\chi}}\frac{\widetilde{L}}%
{\ell_{\text{4D}}^{2}},\label{3Drelation}%
\end{equation}
with $\chi$ a parameter to be fixed by additional physical considerations.

Returning to equation (\ref{deltamsplit}), we obtain an estimate on the mass
splitting:%
\begin{equation}
\Delta m_{\text{3D}}\sim N^{\chi}\frac{\ell_{\text{4D}}^{2}}{L^{2}}\frac
{1}{\widetilde{L}}=\frac{N^{\chi}}{N}\frac{1}{\widetilde{L}},
\end{equation}
where in the last equality we used equation (\ref{WZWradius}).

But this is not the end of the story, because we need to return to the 4D
world. To get there, we simply let $N$ get even bigger. In this limit, the physics is better
described in 4D terms. To get there, we use the relation $\widetilde{L}%
\sim\ell_{\text{4D}}/\sqrt{N}$. So, we get a 4D mass splitting given by:%
\begin{equation}
\Delta m_{\text{4D}}\ \sim\frac{N^{\chi}}{\sqrt{N}}\frac{1}{\ell_{\text{4D}}}.
\end{equation}
We can also relate this to an expression involving the IR\ length scale, via
equation (\ref{IR}):%
\begin{equation}
\Delta m_{\text{4D}}\ \sim\frac{N^{\chi}}{N^{1/4}}\frac{1}{\ell_{\text{IR}}}.
\end{equation}

So far, we have not determined $\chi$. We expect that $\Delta m_{\text{4D}}$ is parametrically
separated from the IR and UV cutoffs in the large $N$ limit. This is really
just a consistency condition that we can make sense of the 4D\ effective field
theory at all:
\begin{equation}
M_{\text{IR}}\ll\Delta m_{\text{4D}}\ll M_{\text{UV}}\text{,}%
\end{equation}
where:%
\begin{equation}
M_{\text{IR}}=\frac{1}{\ell_{\text{IR}}}\text{ \ \ and \ \ }M_{\text{UV}%
}=\frac{1}{\ell_{\text{4D}}}\text{.}%
\end{equation}
Said differently, if we hold fixed these cutoffs, we expect $\Delta m_{\text{4D}}$
to remain fixed. This in turn means we should set the parameter $\chi$ so that
the mass splitting is the geometric mean:
\begin{equation}
\Delta m_{\text{4D}}\sim\sqrt{M_{\text{IR}}M_{\text{UV}}}\text{,}%
\end{equation}
which fixes the parameter $\chi=3/8$ so that:%
\begin{equation}
\Delta m_{\text{4D}}\ \sim\frac{1}{N^{1/8}}\frac{1}{\ell_{\text{4D}}}\sim
N^{1/8}\frac{1}{\ell_{\text{IR}}}.
\end{equation}
It would of course be desirable to perform a first principles calculation of
this exponent.

Summarizing, 3D $\mathcal{N}=1$ supersymmetry ensures the zero point energies
of the 4D system cancel off, but the apparent superpartners of any QFT\ sector
generated by a configuration of intersecting 7-branes will have a split
spectrum which is up by a factor of $N^{1/8}$ relative to the IR\ scale.

Note that in the F-theory models considered in previous sections, the
appearance of gauge groups and matter in complex representations means that
the matter content generated by the \textquotedblleft$Spin(7)$ Standard
Model\textquotedblright\ will have all of the superpartners of the standard
MSSM. What is far less clear is the mass spectrum and interactions, which
could in principle be far more general than what is available in the MSSM with
soft supersymmetry breaking parameters. For example, the absence of holomorphy
constraints / non-renormalization theorems for the superpotential means that
we must permit more gauge invariant interaction terms.

What value of $N$ matches observation? The analysis presented here has assumed
we have a static $S^{3}$, not an expanding one. With this in mind, what we
take for the  IR cutoff is somewhat subtle.

Even so, it seems reasonable to assume that the expansion is \textquotedblleft
slow enough\textquotedblright\ to use the above formulae to extract a
quantitative scale. This in some sense sets a lower bound for the mass
splitting. With these caveats in mind, we treat the present expansion of the
Universe as \textquotedblleft static enough,\textquotedblright\ so we take
$M_{\text{UV}}\sim10^{19}$ GeV and $M_{\text{IR}}\ \sim10^{-12}$ GeV (as set
by the observed dark energy density). The geometric mean then tells us the 4D
mass splitting:\footnote{This numerology has been observed by many
individuals, but usually as a way to motivate the size of the cosmological
constant, not so much as a way to motivate TeV scale supersymmetry.}%
\begin{equation}
\Delta m_{\text{4D}}\sim10^{3.5}\text{ GeV,}%
\end{equation}
which is of order the TeV scale! So at least in this class of models, there is
a compelling reason for the cosmological constant to be quite small, but to
also have O(TeV) scale superpartners. This is clearly a rough order of
magnitude estimate,\footnote{One could imagine a more precise estimate
producing a larger number such as $50$ TeV.} but it does provide an
alternative motivation rather than just naturalness considerations / stability
of the electroweak scale.

On the other hand, the proper IR\ cutoff to take might be higher, as set by
the initial conditions necessary to generate the present day Universe. Then,
the mass of the superpartners would clearly be bigger, and may remain (at
least for now) out of experimental reach. Turning the discussion around,
observing superpartners at a particular mass scale would provide an additional
input on the proper choice of the IR\ cutoff, and the parameter $N$.

We find this line of development promising.

\section{Conclusions and Future Directions}\label{sec:CONC}

In this paper we have laid the groundwork for the study of 4D cosmological
scenarios which arise from F-theory compactified on $Spin(7)$ manifolds. We
have also clarified a number of aspects connected with the 3D physics
associated with M-theory compactified on $Spin(7)$ manifolds. An important
feature of our analysis is the compactification of F-theory on spacetime with
a WZW\ factor. This in particular makes it possible to see why there is no
$Spin(7)$ compactification to flat space, but there is clearly a way to
achieve this on curved backgrounds. Another crucial element of this
description is the interpretation of the 11D supergravity limit as the
untwisted sector of the T-dual orbifold CFT\ description of a WZW model. This
in particular shows how F-theory serves as a UV\ completion for the M-theory
construction. In the remainder of this section we discuss a number of
directions which would clearly be exciting to develop further.

At the level of formal developments, we have seen the appearance of a new sort
of spectral cover construction for the localized degrees of freedom associated
with an F-theory compactification. It would be very interesting to fully catalog the
possible differences with the spectral cover used in much of the earlier F-theory GUTs
literature (see e.g. \cite{Beasley:2008dc, Donagi:2008ca, Hayashi:2009ge, Donagi:2009ra}).
In particular, the analysis of the present paper provides a
helpful starting point for analyzing the local geometry of $Spin(7)$ manifolds
with singular ADE\ fibrations.

As mentioned in the Introduction, one of the motivations for the present work
was to develop explicit string compactifications with a dark energy sector.
Clearly, we have obtained some toy models which accomplish this, and so it
would be interesting to check whether these models can serve as the starting
point for more realistic cosmological scenarios. This can likely be accommodated because
localized QFT sectors on 7-branes and 3-branes
contribute additional matter and radiation sectors.

At a more formal level, it is tempting to view the dark energy dominated
limit of our model as a regulated version of de Sitter space. Since we also have a direct
account of microstates in the WZW model Hilbert space, this suggests a potential avenue of attack
for determining the microscopic origin of the Gibbons--Hawking de Sitter entropy.

The appearance of an instability in our model (much as in the Einstein
static Universe) also suggests a new way to construct inflationary models.
Another interesting feature is that we can clearly
see the appearance of both \textquotedblleft up-tunnelling\textquotedblright%
\ and \textquotedblleft down-tunneling\textquotedblright\ to different
cosmologies by jumps in the total number of $H$-flux units threading the initial
$S^{3}$. In stringy terms, this tunneling comes from Euclidean five-branes wrapped
over the internal six-manifold. For the purposes of inflationary cosmology,
this again is suggestive, because there is a natural way to end
inflation:\ simply tunnel out to a different value of the $H$-flux. This may also
have observational consequences in gravitational wave experiments. It would be interesting
to pursue this further.

It is quite likely that this class of cosmological scenarios has distinctive
phenomenological signatures which could potentially be measured. For example, in
the WZW\ model there is a cutoff on the angular momentum of objects
which scales as $N$ (see also \cite{Heckman:2014xha}). This in turn suggests a corresponding
cutoff on the angular resolution of the observed Cosmic Microwave Background.
Here, we should note that although the present size of
the Universe is clearly much larger than the Planck length, what really enters
in this angular momentum cutoff is the value of the $S^{3}$ prior to the onset
of instabilities leading to rapid expansion in the scale factor.
A related point is that the presence of a three-form flux also
suggests a covariant non-commutative deformation
of 4D physics (see e.g. \cite{Alekseev:1999bs} as well as \cite{Heckman:2014xha}).

Finally, it is of course exciting to see the appearance of a 4D\ model with
``$\mathcal{N}=1/2$ supersymmetry,'' and a clear connection to a 3D $\mathcal{N}%
=1$ system. This appears to imply that we have found a 4D\ system where the zero
point energy of various quantum fields will indeed drop out, even though
$\mathcal{N}=1$ supersymmetry is absent. It would seem worthwhile
to develop the full formalism of an ``$\mathcal{N} = 1/2$ MSSM,'' and its interplay
with the presence of dark energy. This will also likely alleviate
some cosmological issues connected with compactification moduli.
We leave a full analysis for future work.

\newpage

\section*{Acknowledgements}

We thank F.~Apruzzi, V. Balasubramanian, M. Cveti\v{c}, L. Dolan,
F. Hassler, O. Janssen, J. Khoury, D. Robbins, J. Sakstein, and M.
Trodden for helpful discussions. We also thank M. Del Zotto and J. Sakstein for comments on an earlier
draft. We especially thank T. Rudelius for relaying a communication with E. Witten which
lead to several conceptual improvements. JJH thanks the 2018 Summer Workshop at the
Simons Center for Geometry and Physics for hospitality during part of this
work. The work of JJH, CL and GZ is supported by NSF CAREER grant PHY-1756996.
The work of LL\ is supported by DOE Award DE-SC0013528Y.

\appendix

\section{Killing Spinor Equations for Type IIB Supergravity}\label{app:IIB}

The Killing spinor equations for Type IIB supergravity are
\cite{Schwarz:1983qr,Howe:1983sra}
\begin{equation}\label{eqn:IIBKS}
  \begin{aligned}
    \delta \psi_M &= \mathcal{D}_M \epsilon - \frac{1}{96} \left(
    \Gamma_M^{P_1\cdots P_3}G_{P_1\cdots P_3} - 9
  \Gamma^{P_1P_2}G_{MP_1P_2}\right)\epsilon^c +
  \frac{i}{192}\Gamma^{P_1\cdots P_4}F_{MP_1\cdots P_4}\epsilon = 0 \cr
  \delta \lambda &= i \Gamma^M P_M \epsilon^c +
  \frac{i}{24}\Gamma^{P_1\cdots P_3}G_{P_1\cdots P_3}\epsilon = 0 \,.
  \end{aligned}
\end{equation}
Here we have defined
\begin{equation}
  G = i e^{\phi/2}\left(\tau dB_2 - d C_2\right) \,,
\end{equation}
as the $SL(2,\mathbb{Z})$ covariant three-form flux, and further
\begin{equation}
  P = \frac{i}{2\tau_2} d \tau \,.
\end{equation}
The covariant derivative is defined as
\begin{equation}
  \mathcal{D}_M = \nabla_M - i q Q_M \,,
\end{equation}
where $q$ is the charge of the acted on field under the $U(1)_D$ symmetry of
Type IIB \cite{Gaberdiel:1998ui}, and $Q$ is the connection for that abelian symmetry defined as the
variation of the axio-dilaton
\begin{equation}
  Q = - \frac{1}{2\tau_2}d \tau_1 \,.
\end{equation}
The spinor appearing in (\ref{eqn:IIBKS}) is defined as
\begin{equation}
  \epsilon = \epsilon_1 + i \epsilon_2 \,,
\end{equation}
where $\epsilon_i$ are the two Majorana--Weyl supersymmetries of the Type IIB
supergravity, and thus we also immediately see that
\begin{equation}
  \epsilon^c = \epsilon_1 - i \epsilon_2 \,.
\end{equation}
We wish to consider Type IIB on a spacetime of the form
\begin{equation}
  \mathbb{R}_{\text{time}} \times M_3 \times B_6 \,,
\end{equation}
where $B_6$ is a six-manifold that forms the base of a torus fibered
Spin$(7)$ eight-manifold, and generally we will have $M_3 = S^3$. The metric on
the 10D spacetime is the warped product
\begin{equation}
  ds^2 = e^{2C}\left( - dt^2 + a(t)^2 d \Omega^2_{S^3} + ds^2_{B}\right)
  \,,
\end{equation}
where $C \in \Omega^0(B, \mathbb{R})$ is a warp factor. We pick the frame
\begin{equation}
  e^0_{\phantom{0}t} = e^C \,, \quad e^1_{\phantom{1}\psi} = e^C
  a(t) \,, \quad e^2_{\phantom{2}\theta} =
  e^C a(t) \sin(\psi) \,, \quad e^3_{\phantom{3}\phi} = e^C a(t)
   \sin(\psi)\sin(\theta) \,, \quad e^i_{\phantom{i}m} \,,
\end{equation}
where $i$, $m$ are, respectively, the frame and curved indices on $B$, and all
others vanishing. The spin connection has non-zero mixed external-internal
components, which are
\begin{equation}
  \omega_\mu^{\phantom{\mu}ai} = e^a_{\phantom{a}\mu} e^{im}\partial_m C
  \,.
\end{equation}
where we write only the upper triangular entries, with the lower determined by
anti-symmetry of $\omega$, and further we have ignored all terms proportional
the derivative of $a(t)$ as we do not expect a supersymmetric solution away
from a fixed value of $a$.

Let us first consider the the vanishing of the dilatino variation. We want the
three-form flux to be solely and isotropically supported on the $S^3$ factor of the geometry, and
thus it must be proportional to the volume form
\begin{equation}
  G_{abc} = G \epsilon_{abc} \,,
\end{equation}
where lowercase early-alphabet latin indices are frame indices on the $S^3$,
and we take $G$ to be real.  Furthermore, since the torus fibration is
non-trivial only over the internal space, $B$, and thus completing it to an
auxiliary $Spin(7)$ eight-manifold, we have that $P_M$ is supported only over the
$B$ directions. We will also choose to consider only solutions without
fiveform flux, thus $F = 0$. The dilatino equation is that
\begin{equation}
  \Gamma^i P_i \epsilon^c + \frac{1}{4} G \Gamma^{123} \epsilon = 0 \,,
\end{equation}
This relation breaks half of the supersymmetry. One solution is the near
horizon geometry of the NS5-brane \cite{Callan:1991at}, where $P$ is turned on only in
one of the internal directions, which has the form $B = \mathbb{R}^5 \times
\mathbb{R}_{\varphi}$. Next we can study the time-like variation of the
gravitino, where the spinor is chosen not to vary, and one finds
\begin{equation}
  \begin{aligned}
    \frac{1}{4}\omega_t^{AB}\Gamma_{AB}\epsilon - \frac{1}{96} e^C
    \Gamma_{0}^{\phantom{0}bcd}G_{bcd}\epsilon^c &= 0 \cr
    \implies e^{im}\left(\partial_mC\right)\Gamma_i \epsilon -
    \frac{1}{4}G \Gamma^{123}\epsilon^c &= 0 \,.
  \end{aligned}
\end{equation}
Given an appropriate relation among $G$, $\epsilon$, and the axio-dilaton
profile $P$, one can show that this is equivalent to the dilatino variation,
and thus does not break any further supersymmetry, as in the NS5 near horizon
geometry solution.

We are specifically interested in the resulting 4D Killing spinor equations,
on $\mathbb{R} \times S^3$, as the F-theory intuition will guarantee that
exactly one component of the internal spinor on $B$ will be preserved, because
of the $Spin(7)$ holonomy of the auxiliary torus fibered eight-manifold. The spinors
decompose as
\begin{equation}
  \begin{aligned}
    Spin(1,9) &\rightarrow Spin(1,3) \times Spin(6) \rightarrow Spin(3) \times
    Spin(6) \cr
    {\bf 16} &\rightarrow ({\bf 2,1,4}) \oplus ({\bf 1,2,\overline{4}})
    \rightarrow ({\bf 2,4}) \oplus ({\bf 2,\overline{4}}) \,.
  \end{aligned}
\end{equation}
The reality condition on the ${\bf 16}$ becomes that condition that the $({\bf
2,4})$ and the $({\bf 2,\overline{4}})$ are conjugates. We can write
\begin{equation}
  \epsilon_i = \eta_i + \eta_i^c \,,
\end{equation}
where $\eta_i$ are in the $({\bf 2,4})$.
Let us now decompose these $\eta_i$ into spinors on the $S^3$ and the base
$B$. We write
\begin{equation}
  \eta_i = \rho_i \otimes \chi_i \,, \quad \eta_i^c = \rho_i \otimes \chi_i^c
  \,.
\end{equation}
We also decompose the 10D $\Gamma$-matrices as
\begin{equation}
  \Gamma_a = \gamma_a \otimes 1_6 \,, \quad \Gamma_i = \gamma_4 \otimes
  \widetilde{\gamma}_i \,,
\end{equation}
where $\gamma_a$ are the 4D $\Gamma$-matrices (with $\gamma_4$ the 4D
chirality matrix) and $\widetilde{\gamma}_i$ are the 6D $\Gamma$-matrices.

Because the $Spin(7)$ will only preserve one component of either $\chi_1$ or
$\chi_2$ we can consider $\rho_2 = 0$ and $\rho_1 = \rho$. The 4D Killing
spinor equation in the time direction is then
\begin{equation}\label{eqn:KStime}
    \sum_i e_i^{\phantom{i}m}\partial_m C \rho - \frac{1}{8} G \gamma^{123}
    \rho^c = 0 \,.
\end{equation}
The solution to this equation uses the three-form flux to provide a reality
condition that relates $\rho$ to its conjugate. In signature $(3,1)$ this forces the Killing
spinor to vanish, but in signature $(2,2)$ it permits a Majorana--Weyl spinor.

\section{Global Constructions of \boldmath{$Spin(7)$} Manifolds}\label{app:spin7_construction}

In this Appendix, we collect some of aspects of constructing compact $Spin(7)$ manifolds.
Although they have appeared already in Berger's classification \cite{MR0079806} of special holonomy manifolds, the first local example with a $Spin(7)$ metric appeared much later in \cite{MR916718}.
While more local examples with complete $Spin(7)$ metrics have been produced since, the tools to construct global, i.e., compact $Spin(7)$ manifolds have been very limited.
For a long time, the only examples were based on work of Joyce \cite{Joyce1996, Joyce:1999nk}, which was generalized in \cite{2013arXiv1309.5027K}.
So far, the only other type of known construction is motivated by the twisted connected sum (TCS) construction of $G_2$ manifolds \cite{MR2024648, Corti:2012kd, MR3109862, Halverson:2014tya, Halverson:2015vta, Braun:2016igl, Braun:2017ryx, Guio:2017zfn, Braun:2017uku, Braun:2017csz, Braun:2018fdp} and has been dubbed generalized connected sums (GCS) \cite{Braun:2018joh}.

\subsubsection*{Joyce's Quotient Construction}

The simplest construction is an orbifold $T^8 / \Gamma$ \cite{Joyce1996}, where $\Gamma$ is discrete subgroup of the $T^8$ isometry.
A very concrete model of these type has also been presented in \cite{Lee:2014wma}.
In general, the resulting quotient is singular, and Joyce classified those $\Gamma$ which allow for a resolution of $T^8 / \Gamma$ that is compatible with the $Spin(7)$ structure.
Since at generic points, the quotient will still have the structure of products of torus, it is conveivable to have some torus fibration structures within the quotient.
However, it is also clear that, just starting with a direct product of tori, the resulting fibration structures might be limited.

The key observation in \cite{Joyce:1999nk} is that one can also obtain $Spin(7)$ holonomy via anti-holomorphic involutions $\sigma$ of a Calabi--Yau fourfold $Z$ with Ricci-flat metric $g_Z$.
To be more precise, let us denote the holomorphic four-form and the K\"ahler form on $Z$ by $\Omega_{(4,0)}$ and $J$, respectively.
Then the four-form
\begin{equation}
	\Omega_{(4)} := \text{Re} \left( e^{i \theta} \, \Omega_{(4,0)} \right) + \frac{1}{2} \, J \wedge J
\end{equation}
defines a $Spin(7)$ structure on $Z$.
Now suppose there is an anti-holomorphic involution $\sigma: Z \rightarrow Z$, i.e., $\sigma$ satisfies
\begin{align}
	\sigma^2 = \text{id}_Z \, , \quad \sigma^* (g_Z) = g_Z \, , \quad \sigma^*(I) = -I \, ,
\end{align}
where $I$ is the complex structure of $Z$.
These conditions imply that $\sigma^*(J) = -J$ and $\sigma^* (\Omega_{(4,0)}) = e^{2 i \theta} \, \overline{\Omega}_{(0,4)}$ for some $\theta \in [0, 2\pi)$, which can be easily verified to imply the $\sigma$-invariance of $\Omega_{(4)}$ as defined in \eqref{eq:Cayley_fourform_from_CY4}.
Therefore, the quotient $Y_{8} = Z/ \sigma$ inherits the $Spin(7)$ structure $\Omega_{(4)}$, while the Calabi--Yau structure is lost due to the anti-holomorphic involution.

In principle, the anti-holomorphic quotient can also be applied to elliptically fibered Calabi--Yau fourfolds $\pi: Z \rightarrow B$.
In particular, if the involution $\sigma$ respects the fibration structure, i.e., there is an induced involution $\sigma_B : B \rightarrow B$ such that $\pi \circ \sigma \circ \pi^{-1} = \sigma_B$, then the quotient $Y_{8} = Z / \sigma$ has a fibration structure over $\tilde B = B / \sigma_B$ whose generic fiber is again a torus.
However, because of the involution on both fiber and base, the resulting torus fibration $\tilde\pi : Y_{8} \rightarrow \tilde{B}$ no longer has a globally holomorphically varying $\tau$ function \cite{Bonetti:2013fma, Bonetti:2013nka}.
In fact, this can be seen from the various singular fibers that can appear, e.g., a Klein-bottle.
From a Type IIB perspective, where $\tau$ is identified with the axio-dilaton, the non-holomorphic variation can be interpreted as the backreacted background of including NS5-branes, which in turn reduces the number of preserved supersymmetry generators.

For completeness, we note there are certain subtleties about the singularities one obtains from an anti-holomorphic quotient.
The original work of Joyce \cite{Joyce:1999nk} considered only certain involutions $\sigma$ with isolated fixed points.
In these cases, the quotients have a resolution compatible with the $Spin(7)$ structure.
For general involutions the quotient can also have higher dimensional singularities.
However it is currently unclear if, generically, these singularities can also be resolved in an $Spin(7)$ compatible way (see below for case where this is the case).
Nevertheless, as F- and M-theory can also be defined on Calabi--Yau spaces with terminal, i.e., non-crepantly resolvable singularities \cite{Arras:2016evy,Grassi:2018rva}, it should also be possible to give
a sensible physical interpretation of $Spin(7)$ manifolds with higher dimensional singularities.
Indeed, this is what the local gauge theory of section \ref{sec:MTHEORY} provides.

\subsubsection*{Generalized Connected Sums}

A second type of global models has been recently presented within the context of the so-called generalized connected sum (GCS) construction \cite{Braun:2018joh}.
Following the same idea as in the TCS constructions of $G_2$ manifolds, one obtains the $Spin(7)$ manifold by gluing together various building blocks carrying some form of holomorphic / Calabi--Yau structure.
To be more precise, the GCS construction has two building blocks.
One is an open Calabi--Yau fourfold $Z_+ \cong \text{CY}_4$ with an asymptotic cylindrical region where the geometry looks like a Calabi--Yau threefold $\text{CY}_3$ times a cylinder.
The other building block is an open $G_2$ manifold $N_{7}$ times a circle, $Z_- \cong N_{7} \times S^1$, where the $G_2$ manifold has an asymptotic geometry $\text{CY}_3 \times \text{interval}$.
The two blocks are then glued along the cylinder $\cong S^1 \times \text{interval}$.
Note that one can imagine decomposing the $G_2$ manifold inside $Z_-$ further into the building blocks of an TCS construction, which contain K3-fibrations, thus making the previously mentioned holomorphic / CY structure apparent in both parts of the GCS $Spin(7)$ manifold.
In this construction, the Cayley-form can be identified on each of the two building blocks as follows:
On $Z_+ \cong \text{CY}_4$ with holomorphic four-form $\Omega_{(4,0)}$ and K\"ahler form $J$, we simply have $\Omega^+_{(4)} = \text{Re} \, \Omega_{(4,0)} + \frac{1}{2} J \wedge J$, which we already know from Joyce's work to be $Spin(7)$ compatible.
On $Z_- \cong N_{7} \times S^1$, we can construct the four-form $\Omega^-_{(4)} = d \theta \wedge \varphi + \star \varphi$, where $\theta$ is the coordinate on $S^1$ and $\varphi$ the three-form specifying the $G_2$ structure on $N_{7}$.
We refer to \cite{Braun:2018joh} for details of the gluing procedure which identifies $\Omega^+_{(4)}$ with $\Omega^-_{(4)}$ in the asymptotic region.

In \cite{Braun:2018joh}, the authors also presented how the GCS construction relates to anti-holomorphic quotients.
Loosely speaking, a suitable quotient of a CY-fourfold results in a $Spin(7)$ manifold, which locally is a CY$_3$ fibration over a disc.
This part constitutes the $Z_+$ building block of the GCS.
Near the boundary circle, where the local geometry is CY$_3 \times \text{interval} \times S^1$, the involution acts non-trivially on CY$_3 \times \text{interval}$, producing a local $G_2 \times S^1$ geometry \cite{Harvey:1999as}, i.e., the $Z_-$ part of the GCS.
This is also an example where the higher dimensional singularities of the anti-holomorphic involution have a $Spin(7)$ resolution:
Because the involution acts on the CY$_3 \times \text{interval}$ part, any resulting fixed loci will be accompanied by the $S^1$, i.e., have real dimension at least 1.
As explained in \cite{Braun:2018joh}, the resolution of this type of singularities could be related to the appearance of the local $G_2$ geometry, and not an asymptotically locally Euclidean (ALE) space, as it was the case for the resolution of point-like singularities in \cite{Joyce:1999nk}.

Again, in this type of constructions, it is straightforward to have an overall torus fibration structure in the GCS manifold $Y_{8}$.
In particular, given that the TCS construction of $G_2$ manifolds have inherent K3-building blocks, it is plausible that, with the suitable choice of $Z_+ \cong \text{CY}_4$, the resulting $Spin(7)$ space $Y_{8}$ can be at least locally be viewed as a K3-fibration over a four-manifold.
An easy example of this sort is again the Joyce orbifold \cite{Braun:2018joh}.

\section{$\mathcal{N}=2$ Super Quantum Mechanics \label{app:SQM}}

In this Appendix we review some facts about $\mathcal N=2$ super quantum mechanics. The central interest for us will be the different supermultiplets that can be used to build an effective theory. For theories with $\mathcal N=2$ supersymmetry in 1D there are two supermultiplets of interest conventionally called 2A and 2B multiplets. The 2A multiplet descends from (1,1) supersymmetry in 2D and its off-shell degrees of freedom are a real scalar, a complex fermion and a real auxiliary field. The 2B multiplet comes from (0,2) supersymmetry in 2D and its off-shell degrees of freedom are a complex scalar and a complex fermion. While other multiplets are possible they do not come from standard toroidal compactifications and we will not need them in the following. For a more detailed discussion of $\mathcal N=2$ superspace  we refer to \cite{Haupt:2008nu} and references therein.

In the following we will discuss the effective field theory which can be built using 2A multiplets as these will be the ones of interest in section \ref{sec:CLOSED}. It is possible to package all the components of the 2A multiplets in superfields whose expansion in terms of superspace coordinates is
\begin{align}
\Phi = \phi + i \theta \psi + i \bar \theta \bar \psi +\frac{1}{2} \theta \bar \theta F\,.
\end{align}
The effective action can be written as the integral over superspace in terms of a moduli space metric $G_{ij}$ and a superpotential $\mathcal W$. Note that since 2A multiplets are real multiplets both the metric and the superpotential are real functions as opposed to the usual 4D case where the latter is a holomorphic function. The superspace action is
\begin{align}
S_\text{2A} = \frac{1}{4} \int d t \, d^2\theta \left[G_{ij}(\Phi) D \Phi^i \bar D \Phi^j  +\mathcal W(\Phi) \right]\,.
\end{align}
Here $D = \partial_\theta + \frac{i}{2} \bar \theta \partial_t$ and $\bar D = -\partial_{\bar \theta}- \frac{i}{2}  \theta \partial_t$ are superspace covariant derivative. Other terms can be added in the action with different contractions of the superspace covariant derivatives but they will not be of interest for us. After integrating over superspace and eliminating the auxiliary fields via their equations of motion the bosonic components of the action are
\begin{align}
S_{2A} = \frac{1}{16} \int dt \left[ G_{ij} (\phi) \dot \phi^i \dot \phi^j - G^{ij}(\phi) \mathcal W_i(\phi) \mathcal W_j (\phi)\right]\,.
\end{align}
The discussion thus far has employed rigid $\mathcal N=2$ superspace in 1D, however introducing 1D gravity has only a very mild effect. Using the lapse function $\mathcal N(t)$ whose square is the component of the 1D metric tensor it is possible to write the action as
\begin{align}
S^{\text{bosonic}}_{2A} = \frac{1}{16} \int dt \left[ \frac{1}{\mathcal N} G_{ij} (\phi) \dot \phi^i \dot \phi^j - \mathcal N G^{ij}(\phi) \mathcal W_i(\phi) \mathcal W_j (\phi)\right]\,.
\end{align}
While in general there are no constraints on the form of the metric $G_{ij}(\phi)$ we will borrow some conclusions drawn from the case of 11D supergravity on Calabi--Yau fivefolds \cite{Haupt:2008nu}. Of particular interest for us are moduli coming from the K\"ahler form of the internal space for they correspond to 2A multiplets in the 1D action. The metric on the K\"ahler moduli space can be written in terms of a function $\mathcal K$ which can be identified with a K\"ahler potential.\footnote{Note however that the K\"ahler moduli space is not a K\"ahler manifold.} Defining $\kappa = 5! \text{Vol}$ where $\text{Vol}$ is the volume of the compactification space one gets that the metric is
\begin{align}
G_{ij} &= \frac{8}{5!} \kappa \left[\partial_i \partial_j \mathcal K-25 \frac{\kappa_i \kappa_j}{\kappa^2}\right]\,,\\
\mathcal K &= -\frac{1}{2} \log \kappa\,.
\end{align}
Here we take $5\kappa_i = \partial_i \kappa$. The most relevant feature
of this metric is that one of its eigenvalues is negative, a
point which figures prominently in the analysis of section \ref{sec:CLOSED}.

\newpage

\bibliographystyle{utphys}
\bibliography{FDark}

\end{document}